\documentclass[aps,twocolumn,superscriptaddress]{revtex4} 
\usepackage{graphicx}  
\usepackage{dcolumn}   
\usepackage{bm}        
\usepackage{amssymb}   
\usepackage{amsmath}
\usepackage{version}
\usepackage{textgreek}
\usepackage{braket}
\usepackage{setspace} 
\usepackage[colorlinks=true,linkcolor=blue, citecolor=blue]{hyperref}
\usepackage{titlesec}
\newcommand{\be}{\begin{equation}}
\newcommand{\ee}{\end{equation}}
\newcommand{\bea}{\begin{eqnarray}}
\newcommand{\eea}{\end{eqnarray}}
\newcommand{\Eq}[1]{Eq.\,(\ref{#1})}
\newcommand{\Fig}[1]{Fig.\,\ref{#1}}
\newcommand{\Sec}[1]{Sec.\,\ref{#1}}
\newcommand{\Onlinecite}[1]{Ref.\,[\onlinecite{#1}]} 

\newcommand{\Ex}{E_{\rm X}}
\newcommand{\Elp}{E_{\rm LP}}
\newcommand{\Vp}{V_{\rm P}}
\newcommand{\Cx}{c_{\rm X}}
\newcommand{\glp}{\Gamma_{\rm LP}}
\newcommand{\glpem}{\gamma_{\rm LP}}
\newcommand{\glpemmin}{\gamma_{\rm LP}^{\rm min}}
\newcommand{\gsim}{\gamma}

\newcommand{\Rabi}{\Omega_{\rm R}}
\newcommand{\gx}{\gamma_{X}}

\newcommand{\gtwo}{g^{(2)}}
\newcommand{\Df}{{\Delta}_{\rm F}}

\newcommand{\nr}{n_{\rm r}}
\newcommand{\linb}{\hat{\mathcal{L}}}
\newcommand{\annh}{\hat{b}}
\newcommand{\creat}{\hat{b}^{\dagger}}

\newcommand{\annhs}{\hat{b}^2}
\newcommand{\creats}{\hat{b}^{\dagger 2}}
\newcommand{\annht}{\hat{b}^3}
\newcommand{\creatt}{\hat{b}^{\dagger 3}}

\newcommand{\gr}{\gamma_{\rm r}}
\newcommand{\Ham}{\hat{\mathcal{H}}}
\newcommand{\linbp}{\hat{\mathcal{L}}_{\rm p}}
\newcommand{\density}{\hat{\rho}}
\newcommand{\gd}{\gamma_{\rm D}}
\newcommand{\dnr}{\dot{n}_{\rm r}}
\newcommand{\Knj}{K^{\rm nj}}
\newcommand{\Knjd}{K^{\rm nj^{\dagger}}}
\newcommand{\Kndet}{K^{\rm det}_{n}}
\newcommand{\Kndetd}{K^{\rm det^{\dagger}}_{n}}
\newcommand{\Knndet}{K^{\rm nd}_{n}}
\newcommand{\Knndetd}{K^{\rm nd^{ \dagger}}_{n}}
\newcommand{\Knp}{K^{\rm p}_n}
\newcommand{\Knpd}{K^{\rm p^{\dagger}}_n}
\newcommand{\freqF}{\omega_{\rm F}}
\newcommand{\gF}{\gamma_{\rm F}}
\newcommand{\nrav}{\bar{n}_{\rm r}}

\newcommand{\probocc}{\overline{p}_{n}}
\newcommand{\proboccnr}{\overline{p}_{n, \nr}}

\newcommand{\Ns}{N_{\rm s}}
\newcommand{\Nsfit}{N_{\rm s}^{\rm fit}}
\newcommand{\hc}{h_{\rm c}}
\newcommand{\hu}{h_{\rm u}}
\newcommand{\oex}{\omega_{X}}

\newcommand{\snr}{\sigma_{\rm r}}
\newcommand{\aone}{\alpha_1}
\newcommand{\atwo}{\alpha_2}
\newcommand{\gs}{g_{\rm s}}
\newcommand{\gt}{g_{\rm t}}
\newcommand{\gbx}{g_{\rm PB}}

\newcommand{\gtx}{g_{\rm PT}}
\newcommand{\gprime}{g^{\prime}}
\newcommand{\gpt}{g_{\rm PT}}

\newcommand{\nav}{\overline{n}}
\begin{document}


\title{Probing many-body correlations using quantum-cascade correlation spectroscopy}

\author{Lorenzo Scarpelli}
\affiliation{School of Mathematical and Physical Sciences, Macquarie University, Sydney, New South Wales, Australia.}
\affiliation{ARC Centre of Excellence for Engineered Quantum Systems, Macquarie University, Sydney, New South Wales, Australia.}
\author{Cyril Elouard}
\affiliation{Inria, Centre de Lyon, 69 603 Villeurbanne, France.}
\affiliation{ENS Lyon, LIP, 69364 Lyon, France.}
\author{Mattias Johnsson}
\affiliation{School of Mathematical and Physical Sciences, Macquarie University, Sydney, New South Wales, Australia.}
\affiliation{ARC Centre of Excellence for Engineered Quantum Systems, Macquarie University, Sydney, New South Wales, Australia.}
\author{Martina Morassi}
\affiliation{Universit\'{e} Paris-Saclay, CNRS, Centre de Nanosciences et de Nanotechnologies (C2N), 91120, Palaiseau, France.}
\author{Aristide Lemaitre}
\affiliation{Universit\'{e} Paris-Saclay, CNRS, Centre de Nanosciences et de Nanotechnologies (C2N), 91120, Palaiseau, France.}
\author{Iacopo Carusotto}
\affiliation{INO-CNR BEC Center and Dipartimento di Fisica, Università di Trento, I-38123 Povo, Italy.}
\author{Jacqueline Bloch}
\affiliation{Universit\'{e} Paris-Saclay, CNRS, Centre de Nanosciences et de Nanotechnologies (C2N), 91120, Palaiseau, France.}
\author{Sylvain Ravets}
\affiliation{Universit\'{e} Paris-Saclay, CNRS, Centre de Nanosciences et de Nanotechnologies (C2N), 91120, Palaiseau, France.}
\author{Maxime Richard}
\affiliation{MajuLab, CNRS-UCA-SU-NUS-NTU International Joint Research Unit, 117543 Singapore, Singapore}
\affiliation{Centre for Quantum Technologies, National University of Singapore, 117543 Singapore, Singapore}

\author{Thomas Volz}
\affiliation{School of Mathematical and Physical Sciences, Macquarie University, Sydney, New South Wales, Australia.}
\affiliation{ARC Centre of Excellence for Engineered Quantum Systems, Macquarie University, Sydney, New South Wales, Australia.}
\date{\today}

\begin{abstract}
The radiative quantum cascade, i.e. the consecutive emission of photons from a ladder of energy levels, is of fundamental importance in quantum optics. For example, the two-photon cascaded emission from calcium atoms was used in pioneering experiments to test Bell inequalities\,\cite{AspectPRL82,GrangierEL86}. In solid-state quantum optics, the radiative biexciton-exciton cascade has proven useful to generate entangled-photon pairs\,\cite{AkopianPRL06}. More recently, correlations and entanglement of microwave photons emitted from a two-photon cascaded process were measured using superconducting circuits\,\cite{GasparinettiPRL17}. All these experiments rely on the highly non-linear nature of the underlying energy ladder, enabling direct excitation and probing of specific single-photon transitions. Here, we use exciton polaritons to explore the cascaded emission of photons in the regime where individual transitions of the ladder are not resolved, a regime that has not been addressed so far. We excite a polariton quantum cascade by off-resonant laser excitation and probe the emitted luminescence using a combination of spectral filtering and correlation spectroscopy. Remarkably, the measured photon-photon correlations exhibit a strong dependence on the polariton energy, and therefore on the underlying polaritonic interaction strength, with clear signatures from two- and three-body Feshbach resonances. Our experiment establishes photon-cascade correlation spectroscopy as a highly sensitive tool to provide valuable information about the underlying quantum properties of novel semiconductor materials and we predict its usefulness in view 
of studying many-body quantum phenomena
\end{abstract}

\maketitle

\begin{figure*}[]
	\includegraphics[width=\textwidth]{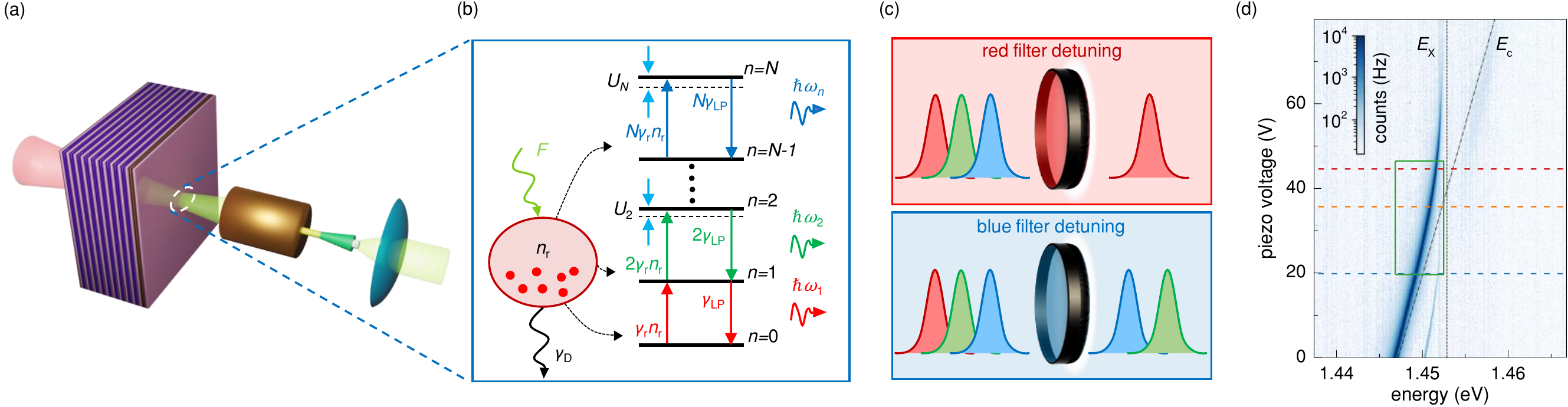}
	\caption{{\bf The polariton quantum cascade and experimental characterization} (a)\,Sketch of the fiber-cavity system. (b)\, Schematic representation of the PQC generation process for a polariton interaction constant $g,\gprime>0$, resulting in positive energy shift $U_{n}$ of the energy levels. An off-resonant pump laser $F$ excites an incoherent long-lived excitonic population, $\nr$, the excitonic reservoir. The reservoir can either decay via dark channels with a rate $\gd$, or induce upward transitions in the polaritonic anharmonic ladder with an effective rate, for the population of the $n^{\rm th}$ level, given by ($n+1$)$\gr \nr$, with $\gr$ the reservoir relaxation rate and $\forall n>0$. Downward transitions from the $n^{\rm th}$ level, with a rate $n\gamma_{\rm LP}$, correspond to the PQC, which leads to multi-coloured photon emission events. We work in the regime $\gr \nr < \gamma$, where the polariton occupation statistics is thermal. (c)\, Illustration of spectral filtering of the photon emission events (Lorentzian wave packets) from the PQC, in the case $g,\gprime>0$. With the spectral filter tuned to the red (top panel), the probability of detecting cascaded photons is reduced, resulting in a suppression of the quantum noise. Vice versa, with the filter tuned to the blue (bottom panel), detection of cascaded photons becomes more likely, therefore increasing the quantum noise. (d)\,White light transmission spectra for different piezo voltages (corresponding to cavity length) revealing strong coupling. $\Ex$ and $E_{\rm c}$ are the exciton energy and the cavity dispersion extracted from the coupled oscillator model, respectively. The blue, orange and red horizontal dashed lines indicate the cavity lengths corresponding to the excitonic contents in the measurements displayed in \Fig{fig:Results}(a,b), \Fig{fig:Results}(c,d) and in the inset of \Fig{fig:g2_resonances}, respectively. Green area indicates the parameter range covered by the measurement results shown in \Fig{fig:g2_resonances}.}
	\label{fig:ExperimentalArrangement}
\end{figure*}
We investigate a semiconductor microcavity in which the strong coupling regime is achieved between cavity photons and excitons, bound electron-hole (e-h) pairs, from a quantum well (QW) placed at the cavity mode antinode. The resulting elementary excitation of the system are  known as exciton-polaritons (polaritons), quasi-particles that behave essentially like photons, except for a few key differences. Owing to their half electronic nature, they can be generated by non-resonant (incoherent) excitations, via electronic relaxation, and Coulomb exchange results in significant interactions between polaritons, and hence to a large effective Kerr-like nonlinearity. In the large polariton number mean field regime, this nonlinearity is at the basis of a quantum fluid behaviour with phenomena such as superfluidity\,\cite{AmoNP09} and hydrodynamic nucleation of solitons and vortices\,\cite{AmoS11,NardinNP11}. In the few polaritons regime, also quantum effects, such as intensity squeezing\,\cite{BoulierNC14}, antibunching\,\cite{MatutanoNMa19, DelteilNMa19} and few-photon phase rotation\,\cite{KuriakoseNP22} have been observed as a result of polariton interactions. Furthermore, polaritonic Feshbach resonances (FRs), i.e., a two-body scattering process with a  biexciton (two bound excitons of opposite spin) state, have been observed using pump-probe spectroscopy\,\cite{TakemuraNPh14}. Even higher-order e-h correlated states\,\cite{TurnerN10} have been found to participate in the polaritonic nonlinearity using multi-wave mixing experiments, albeit in a regime of macroscopic polariton number\,\cite{WenNJOP13}. All these few-body scattering mechanisms determine the polaritonic anharmonic ladder of energy levels, where the energy of each level depends on the number of particles in the system. Contrary to atoms and other standard cavity quantum electrodynamics platforms though, the polaritonic anharmoncity is, up to now, still smaller than the losses in the system, so that the individual transitions of the ladder are not accessible. Remarkably, using our sensitive technique based on quantum-cascaded correlation spectroscopy, we are able to clearly distinguish the contribution of many-body e-h complexes to the polariton interaction, which is an important fact (i) to understand the nontrivial behaviour of the interaction as a function of energy, and (ii) to develop future strategies aimed at enhancing nonlinearities.  

The polariton quantum cascade (PQC) is explored in a fibre-based microcavity system (see \Fig{fig:ExperimentalArrangement}(a)). With the fibre-cavity\,\cite{BesgaPRAp15}, we conveniently engineer a laterally confined cavity mode which results in a tight confinement for the polaritons via the strong coupling regime. This lateral confinement results in a single longitudinal and transverse polariton modes, and in enhanced polariton nonlinearities\,\cite{FinkNMa18, MatutanoNMa19, DelteilNMa19}. The polariton transition frequency $\omega_n$, describing transitions between the $\ket{n}$ and $\ket{n-1}$ states, shifts for increasing number of intra-cavity excitations as
\be
\omega_n=\omega_{\rm LP} + g(n-1) + \gprime(n-1)(n-2) + \dots
\label{eq:omegan}
\ee
where $n$ is the polariton number, $\omega_{\rm LP}$ the $\ket{1}\to \ket{0}$ emission frequency and $g$ ($\gprime$) the 2-polariton (3-polariton) interaction constant. This results in an anharmonic ladder of energy levels which we use to resolve the PQC. \\
We optically excite the system off-resonantly, at a laser power well below condensation threshold to work in the spontaneous emission (SE) regime, where a steady-state population of reservoir excitons is created, and relaxes to generate polaritons (\Fig{fig:ExperimentalArrangement}(b)). Under these conditions, the polariton occupation probability $\probocc$ obeys a black-body cavity statistics\,\cite{ScullyPR67,KlaasPRL18} and the corresponding emitted photons exhibit a bunched zero-delay second-order correlation function $\gtwo(0)=2$. \\
In the presence of a finite nonlinearity due to polariton interactions, the PQC produces multi-coloured photon emissions, with the different photon colours corresponding to different rungs of the PQC. By spectrally filtering the coloured PQC photons with a narrow spectral filter, we can thus modify their (quantum) statistics by an amount that scales like the strength of the polariton interactions. Assuming repulsive interactions, when the filter is red detuned with respect to the polariton SE spectrum, the probability of detecting two-photon cascade emission events from higher transitions in the polariton ladder is reduced (see sketch in \Fig{fig:ExperimentalArrangement}(c)). On the contrary, when the filter is blue detuned, detection of cascade events are more likely. Therefore, by scanning the filter central transmission frequency across the polariton SE spectrum, an ``S-shaped'' modulation of the second order correlation function at zero delay $\gtwo(\tau)$ is expected, with reduced (increased) two-photons correlations with the filter on the red (blue) side of the polariton SE spectrum. The modulation will instead flip sign in case of attractive interactions. Therefore, our method is sensitive to both the magnitude and sign of interactions.\\
We emphasize that in order to be sensitive to few-particle nonlinearities, it is crucial to work in the SE regime where the incoherent emission from few-polariton states dominate the emission dynamics, in contrast to the polariton condensation regime, where the emission statistics is determined by the complex dynamics of the macroscopic coherent condensate\,\cite{SilvaSR16,CarusottoRMP13}.
\begin{figure*}[]
	\includegraphics[width=\textwidth]{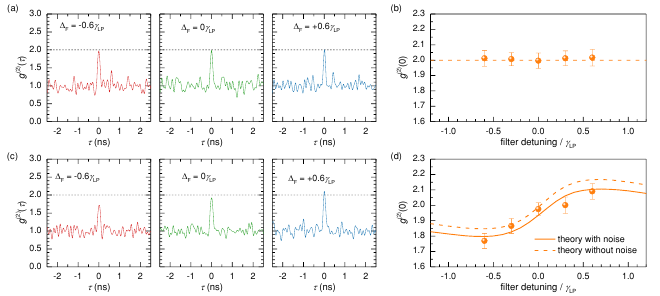}
	\caption{{\bf Correlation spectroscopy as a function of filter detuning} Top panel: (a)\,Measured $\gtwo(\tau)$ for three different filter to polariton peak energy detunings, from red (left) to blue (right) detunings, with the exciton Hopfield coefficient fixed to $|\Cx|^2=0.17$ (blue dashed line in \Fig{fig:ExperimentalArrangement}(d)). (b)\,$\gtwo(0)$ extracted from the fit. (c,d)\,The same as in (a,b), but for an exciton Hopfield coefficient $|\Cx|^2=0.42$ (orange dashed line in \Fig{fig:ExperimentalArrangement}(d)). In this case, the expected dispersive shape is observed. Dashed lines in (b) and (d) are the results of the simulations using \Eq{eq:sim_g2main}. Solid line in (d) is the result of the simulations including the reservoir noise. Details of the theory are described in the Supplement.}
	\label{fig:Results}
\end{figure*}
\begin{figure}[]
	\includegraphics[width=\columnwidth]{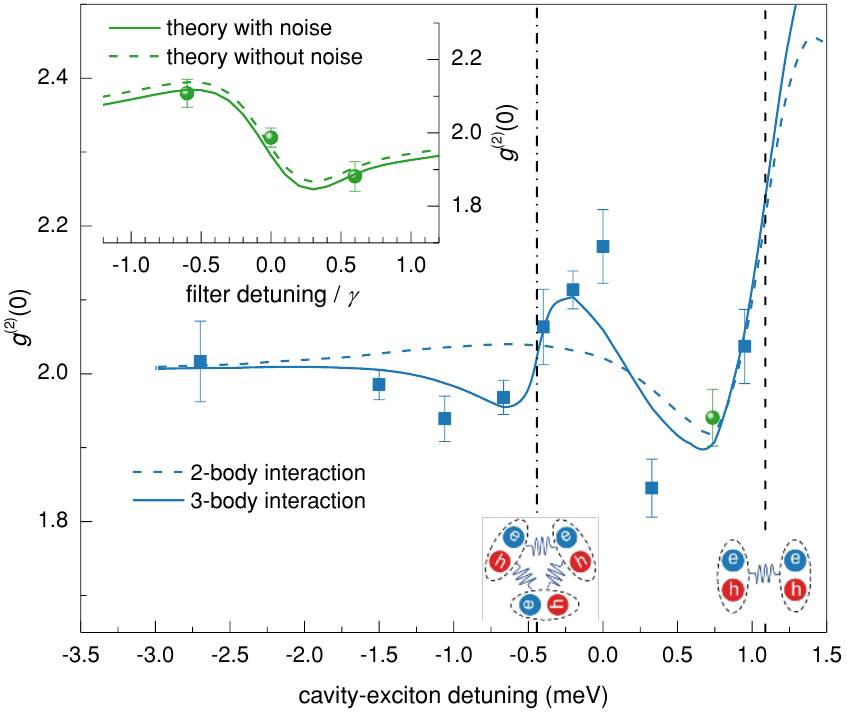}
	\caption{{\bf Signatures of quantum many-body correlations between polaritons} $\gtwo(0)$ as a function of the cavity-exciton detuning with the spectral filter fixed at +0.6$\glpem$. Blue dashed line: model with singlet and triplet polariton interactions, including a FR with the biexciton. Blue solid line: as dashed line but with the additional triexciton resonance. Vertical dash and dash-dot lines: position of the biexciton and triexciton resonances, respectively. Inset: $\gtwo(0)$ versus filter detuning measured at the cavity-exciton detuning corresponding to the green data point and to the red dashed line in \Fig{fig:ExperimentalArrangement}(d). Green solid (dashed) line: model using the nonlinearity obtained from the model including both biexciton and triexciton resonances, with (without) noise.}
	\label{fig:g2_resonances}
\end{figure}

To probe the PQC by correlation spectroscopy, the photons emitted from the cavity and transmitted through a narrow spectral filter are sent to a Hanbury Brown and Twiss interferometer to measure the second order correlation function $\gtwo(\tau)$ as a function of the time delay $\tau$. The optically active material of our microcavity system is an InGaAs QW placed at the center of a $\lambda/2$-cavity in correspondence with an antinode of the TEM$_{00}$ cavity mode. Due to residual birefringence of the GaAs, the polariton lowest-energy mode splits into two orthogonal linearly-polarized modes, separated by about 190 μeV (see Supplement). In all the correlations experiments, we reject the horizontally-polarized emission from the higher lying mode using a combination of waveplates and polarizers, and detect only the vertically-polarized mode. \Fig{fig:ExperimentalArrangement}(d) shows the transmission spectra of a white light probe, generated by a supercontinuum source, as a function of the piezo voltage $\Vp$ controlling the cavity length and therefore changing the caving energy $E_{\rm c}$. The strong coupling between the cavity TEM$_{00}$ mode and the QW exciton is visible as a clear anti-crossing between the lower polariton (LP) and upper polariton (UP) modes. At $\Vp < 30$\,V, we also observe the strong coupling with a higher-order cavity  mode. The data from the TEM$_{00}$ mode are fitted with a coupled oscillator model, yielding a Rabi splitting of 3.0\,meV and an excitonic transition energy $\Ex$=1452.1\,meV. We use this fit to also calculate the exciton Hopfield coefficient $|\Cx|^2$ as a function of $\Vp$, describing the excitonic fraction within the polariton mode.\\
In what follows, we can disregard the UP polariton state as it is well-split from the LP state we are interested in and it is fully rejected by the spectral filter.\\
Additional characterization of the LP mode is performed using high-resolution resonant laser scans (see Supplement), from which we estimate a cavity linewidth of about 64\,$\mu$eV. Importantly, we do not observe any signature of a charged exciton state that behaves as an additional undesirable loss channel for polaritons, which instead was found to be quite large in the sample structure used in \Onlinecite{MatutanoNMa19}. \\
Finally, we perform power-dependent measurements in off-resonant excitation, from which we extract the suitable power range for the SE regime (see Supplement).

We now proceed with a series of correlation measurements. For all these measurements, both excitation power and wavelength are actively stabilised, and the filter linewidth is set to approximately $\glpem/3$, with $\glpem$ the SE linewidth of the vertically polarized LP state. We first fix $|\Cx|^2=0.17$ (corresponding to a cavity-exciton detuning $\Delta=E_{\rm c}-\Ex=-2.7$\,meV), indicated as a blue dashed line in \Fig{fig:ExperimentalArrangement}(d), for which we expect negligible polariton-polariton interactions due to small excitonic fraction within the polariton mode. We measure $\gtwo(\tau)$ for five different filter detunings $\Df$ from the SE peak centre. The noise filtered data (see Supplementary Material for noise filtering procedure) for $\Df=-0.6\glpem\, , \Df=0\glpem$ and $\Df=+0.6\glpem$ are shown in \Fig{fig:Results}(a). In all the measurements, we observe a clear bunching peak around $\tau=0$. The shape of the peak is Gaussian, as determined by the Fourier transform of the filter transmission spectrum, which is also Gaussian (see Supplement). Therefore, we extract the value of the second order correlation function at zero delay $\gtwo(0)$ by fitting $\gtwo(\tau)$ with such a lineshape, from which we also obtain the peak width $\sigma\approx 57$\,ps. Importantly, as the detection response time is $\sigma_{\rm det}\approx 10$\,ps, the deconvolution does not significantly affect the measured correlations, which is another crucial advantage of our method. The extracted values of $\gtwo(0)$ as a function of filter detunings are shown in \Fig{fig:Results}(b), with the error bars representing one standard deviation. For all filter detunings, we measure $\gtwo(0)\approx 2$. This value is consistent with what is expected from the thermal distribution of polaritons that we generate by pumping the system in the SE regime. Furthermore, as the $\gtwo(0)$ does not vary with the filter detuning, it means that the polariton-polariton interaction is indeed so small, as compared to $\glpem$, that the polariton ladder is essentially harmonic. Therefore, by scanning the filter across the SE spectrum, we only change the photon detection rate without modifying the statistics imposed by the polariton occupation. 

We then move on to another cavity detuning where $|\Cx|^2=0.42$ ($\Delta=-0.5$\,meV), corresponding to the orange dashed line in \Fig{fig:ExperimentalArrangement}(d). The noise filtered data are shown in \Fig{fig:Results}(c). In this case, the amplitude of the bunching peak clearly changes with the filter spectral position: our quantitative analysis shows that $g^{(2)}(0)=1.77\pm 0.05$ for $\Df=-0.6\glpem$, i.e. on the red side of the LP resonance, and increases to $g^{(2)}(0)=2.09\pm 0.05$ for $\Df=+0.6\glpem$. The full analysis is shown in \Fig{fig:Results}(d), where a clear ``S-shaped'' modulation is observed. The positive sign of the dispersion implies repulsive interactions. \\
To capture this behaviour more quantitatively, we develop a theoretical framework where the probability $P_{n}^{\rm F}$ of transmitting a photon emitted from the $n^{\rm th}$ transition is modeled as an overlap integral between the filter transmission function $G_{\freqF}^{\gF}(\omega)$, where $G_{\freqF}^{\gF}$ is a Gaussian probability density function with center frequency $\freqF$ and linewdith $\gF$, and the single-photon spectrum $L^{\gsim}_{\omega_n}(\omega)$, where $L^{\gsim}_{\omega_n}$ is a normalized Lorentzian function with center frequency $\omega_n$ and linewidth $\gsim$. The $\gtwo(0)$ can then be calculated as
\be
\gtwo(0) = \frac{\sum_{n} P_{n}^{\rm F} P_{n-1}^{\rm F} n (n-1) \probocc}{\sum_{n} ( P_{n}^{\rm F} n \probocc )^2}\, ,
\label{eq:sim_g2main}
\ee
and it is entirely determined by the probability $P_{n}^{\rm F}$ and the polariton occupation probability $\probocc$.\\
The dashed line in \Fig{fig:Results}(b) and (d) is the result of the model with $g=0$ and $g=0.04\,\glpem \approx 2.7\,\mu$eV respectively, where we assumed $\gprime=0$ for simplicity. In \Fig{fig:Results}(d), in solid line, we also show the model with a modified version of \Eq{eq:sim_g2main}, where we included fluctuations in the excitonic reservoir. This model is in better agreement with the data than the corresponding model without fluctuations, which is consistent with the expected larger interactions between reservoir excitons and polaritons at higher excitonic contents (further details in the Supplement).\\
Importantly, these set of measurements show that, beside being sensitive to few-particle nonlinearities, our method can be used to control and reduce the quantum noise of thermal light. For $g / \gamma \approx 1$, single-photon emitters would be achieved in this experimental setting (see Supplement). This is remarkably different from the conventional quantum blockade strategy\,\cite{VergerPRB06} and other strategies to achieve tunable photon statistics, such as, for example, the unconventional photon blockade\,\cite{LiewPRL10,VanephPRL18,SnijdersPRL18}, spectrally filtered resonance fluorescence\,\cite{HanschkePRL20, PhillipsPRL20} or the Fano effect\,\cite{FosterPRL19}. In these strategies, an input coherent field is needed\,\cite{CasalenguaPRA20}, and the system acts as a nonlinear filter that converts the input Poissonian statistics into sub- or super- Poissonian light. In our configuration, we manipulate the statistics of the direct SE from the system, independently on how it is brought to the excited state. This means that we could electrically inject polaritons and still provide single photons. 

To complete our picture of polariton nonlinearities, we measure correlations along a continuous scan of the excitonic content, covering the region highlighted by the rectangle in \Fig{fig:ExperimentalArrangement}(d). For each value of $\Cx$, we fix the filter detuning to $+0.6\glpem$. The resulting $\gtwo(0)$ as a function of the cavity-exciton detuning is shown in \Fig{fig:g2_resonances}. We obtain a highly non-monotonous behaviour: at large negative detuning, we recover $\gtwo(0)\approx 2$ as expected in absence of nonlinearites. Up to about $\Delta=-1.25$\,meV, $\gtwo(0)$ slightly decreases down to a value of 1.95, to then increase up to 2.2 at $\Delta=0$\,meV. Around this value of $\Delta$, the $\gtwo(0)$ sharply decreases down to a value of about 1.85, to then rapidly increase again. Importantly, the sharp reduction of the $\gtwo(0)$ from a value well above 2 to a value well below 2 corresponds to a sign flip of the interaction from repulsive to attractive.\\
To explain this remarkable behaviour, we recall that the we are probing a linearly polarized polariton mode. In this situation, both the triplet $\aone$ and the singlet $\atwo$ interaction channels between polaritons with parallel or anti-parallel spins respectively, contribute to the measured nonlinearity. Therefore, for the linearly polarized polaritons, the interaction strength to be inserted in \Eq{eq:omegan} reads $g=(\aone+\atwo)/2$, where $\alpha_{1,2}$ are\,\cite{TakemuraPRB17}
\begin{subequations}
\begin{align}
\aone &= \gt |\Cx|^4  \label{eq:alpha1}\\
\atwo &= \gs |\Cx|^4 + 2\gbx^2 |\Cx|^4  \frac{2\Elp - \epsilon_{\rm B}}{\left(2\Elp - \epsilon_{\rm B} \right)^2 + \gamma_{\rm B}^2} \label{eq:alpha2} \,.
\end{align}
\label{eq:alphas}
\end{subequations}
Here $\gt$ and $\gs$ are the triplet and singlet exciton-exciton interactions respectively. The singlet channel contributing to $g$ is of particular interest and central to this work: it is strongly modulated by the presence of correlated excitonic complexes via the FR scattering mechanism. The biexciton, (two bound excitons of opposite spin) contribution is described by the second term in \Eq{eq:alpha2}, with $\gbx$ the polariton-biexction coupling strength, $\epsilon_{\rm B}=2\Ex - E_{\rm B}$ the biexciton energy, $E_{\rm B}$ the biexciton binding energy and $\gamma_{\rm B}$ the biexciton decay rate. For $\Elp < \epsilon_{\rm B}/2$, the FR-induced interaction is attractive, while for $\Elp > \epsilon_{\rm B}/2$ is repulsive. Plugging the polariton nonlinearity obtained with \Eq{eq:alphas} into our analytical theory for the $\gtwo(0)$, and neglecting the reservoir noise for simplicity, we obtain the model shown in \Fig{fig:g2_resonances} in dashed lines. 

While this simple model captures a part of the trend and the sign flip of the interaction, it misses a resonant feature around $\Delta \approx 0\,$meV. According to the literature\,\cite{LevinsenPRL19}, this feature is consistent with the contribution of a three-exciton bound state (a ``triexciton'') to the interactions.\\
To include this possibility, we evaluate the effective interaction terms associated with polariton-biexciton and polariton-triexciton couplings using the adiabatic elimination of the multiexciton complex dynamics (see Supplementary Material). We retrieve the expression for $\alpha_2$ given in \Eq{eq:alpha2}, and derive the three-body interaction contribution $\gprime$ in \Eq{eq:omegan}, which reads:
\be
\gprime = 2\gpt^2 |\Cx|^6  \frac{3\Elp - \epsilon_{\rm T}}{\left(3\Elp-\epsilon_{\rm T} \right)^2 + \gamma_{\rm T}^2}\,, 
\label{eq:alpha3}
\ee
with $\gpt$ the coupling strength with the triexciton, $\epsilon_{\rm T}=3\Ex - E_{\rm T}$ the triexciton energy and $E_{\rm T}$ and $\gamma_{\rm T}$ the triexciton binding energy and decay rate, respectively.\\
The result of the model is shown in \Fig{fig:g2_resonances} in solid lines, where we used $\gs + \gt =6.1\,\mu$eV, $\gbx=0.07$\,meV, $\gtx=0.23$\,meV and $\gamma_{\rm B}=\gamma_{\rm T}=0.34$\,meV. The biexciton and triexciton binding energies are fixed to previously suggested values of $E_{\rm B}=2.2$\,meV\,\cite{TakemuraPRB17} and $E_{\rm T}=2.4E_{\rm B}$\,\cite{LevinsenPRL19}, which result in the resonance energies given by the dash and dash-dot vertical lines, respectively. In this case, the model captures very nicely the overall data trend. Note that our model neglects the inhomogeneous broadening of the biexciton and triexciton resonances, which could further improve the agreement.\\
Most remarkably, for the data point highlighted in green in \Fig{fig:g2_resonances}, which corresponds to the exciton-photon detuning given by the red dashed line in \Fig{fig:ExperimentalArrangement}(d), our measurements suggest an effective attractive interaction. Indeed, as $\gtwo(0)<2$ with the filter on the blue side, we expect a negative dispersion of the $\gtwo(0)$ when scanning the filter detuning at this particular cavity-exciton detuning. In the inset, we show the corresponding measurement, together with the theoretical curves, calculated with and without noise and using the values of $g$ and $\gprime$ at this particular detuning as extracted from the model, showing excellent agreement. Comparing this measurement with the equivalent one for repulsive interactions shown in \Fig{fig:Results}(d), one can clearly see the key difference between overall repulsive or attractive interactions, which gives rise to ``S-shaped'' curves with opposite slopes. These results highlight the sensitivity of our technique to not only the magnitude of interactions, but also to their sign.  

In conclusion, we demonstrate that quantum cascade correlation spectroscopy is an extremely powerful and sensitive technique to investigate anharmonic energy ladders, even when the nonlinearity is smaller than the linewidth and individual transitions are not resolved. \\
Using this technique, we are able to detect polariton scattering processes with two and three-particle bound states, and therefore it allows studying many-body quantum  correlations in a semiconductor.\\
Furthemore, our scheme provides a straightforward way to control and reduce the quantum noise of chaotic light. In the future, as soon as our setup is combined with stronger, yet realistic\,\cite{ChristensenARX22}, nonlinearities with $g / \gamma \approx 1$, single-photon emission events will start to dominate and the emitted light will acquire a markedly quantum character, such as strong antibunching properties. This is of high interest for applications, as our method can be used to achieve electrically-pumped scalable single photon emitters, contrary to other state-of-the-art methods which instead require coherent optical fields as an input. Our technology in principle allows for scaling up the device into arrays of identical single-photon emitters.\\
Finally, our method is applicable to any material system where a radiative cascade can be induced and efficiently detected. For example, in two-dimensional materials, it can serve as a new technique to investigate interactions in novel states of matters, such as incompressible charge states\,\cite{ShimazakiN20} and Wigner crystals\,\cite{SmolenskiN21}. \\

\noindent We would like to thank Guillermo Mu\~{n}oz Matutano and Andrew Wood for early experimental work, and Alexia Auffeves for early contributions towards the theoretical modelling. We also thank Daniele De Bernardis for discussions. We acknowledge financial support from the Australian Research Council Centre of Excellence for Engineered Quantum Systems EQUS (CE170100009). IC acknowledges financial support from the European Union H2020-FETFLAG-2018-2020 project ``PhoQuS'' (n.820392), from the Provincia Autonoma di Trento, and from the Q@TN initiative. C2N acknowledges support from the Paris Ile-de-France Région in the framework of DIM SIRTEQ, the French RENATECH network, the H2020-FETFLAG project PhoQus (820392), the QUANTERA project Interpol (ANR-QUAN-0003-05), and the European Research Council via the project ARQADIA (949730).\\

\noindent {\bf Author contributions} L.S. built the experimental set-up, performed the experiments and analysed the data. C.E. and L.S. realized the theoretical calculations and numerical simulations, with the help of M.J. and M.R.. S.R. and A.L. contributed to the design of the sample structure and M.M. and A.L. grew the sample by molecular beam epitaxy. J.B., S.R. and I.C. participated in the scientific discussions to finalize the work. L.S., C.E., M.R. and T.V. wrote the manuscript, with varying contributions from all authors. L.S., M.R. and T.V. conceived the idea for the experiment. T.V. supervised the project.

\bibliography{OffResonantCorrelations}

\clearpage

\begin{appendix}

\renewcommand{\thefigure}{S\arabic{figure}}	
\setcounter{figure}{0} 

\renewcommand{\theequation}{S\arabic{equation}}	
\setcounter{equation}{0} 

\titleformat{\subsection}{\normalfont\normalsize\itshape\centering}{\thesubsection.}{1em}{}
	
\section*{SUPPLEMENTARY INFORMATION}
\subsection{Fiber cavity characterization} \label{sec:Fiber_Mirror} 
\begin{figure}[]
	\includegraphics[width=0.9\columnwidth]{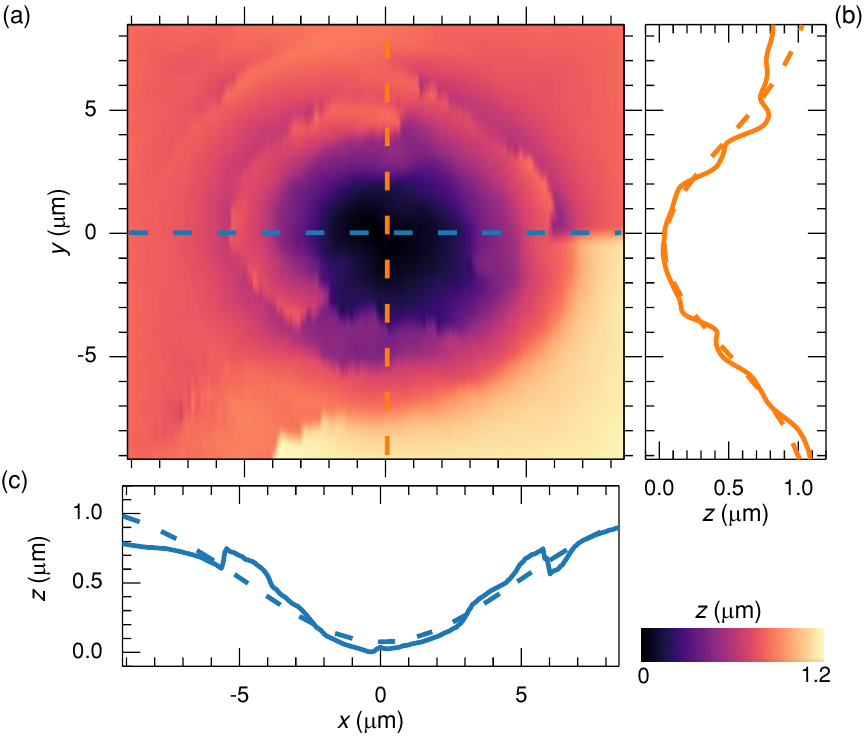}
	\caption{(a)\,Two-dimensional fiber dimple profile obtained from the measured interferogram by two-dimensional phase unwrapping. (b)\,$z$ elevation obtained from the $y$-cut shown in (a) in orange dashed line. (c)\,$z$ elevation obtained from the $x$-cut shown in (a) in blue dashed line. Dashed lines in (b) and (c) are fit to the data.}
	\label{fig:FiberA14_Phase}
\end{figure}

The fiber mirror is a dielectric DBR made of 13.5 pairs of Ta$_{2}$O$_5$/${\rm SiO}_2$, deposited on the fiber tip. Before depositon, the fiber tip was shaped by ${\rm CO}_2$ laser ablation, forming a dimple at the center, whose elevation can be approximated with a two-dimensional Gaussian profile\,\cite{HungerNJOP10}.
\begin{figure}[h]
	\includegraphics[width=0.9\columnwidth]{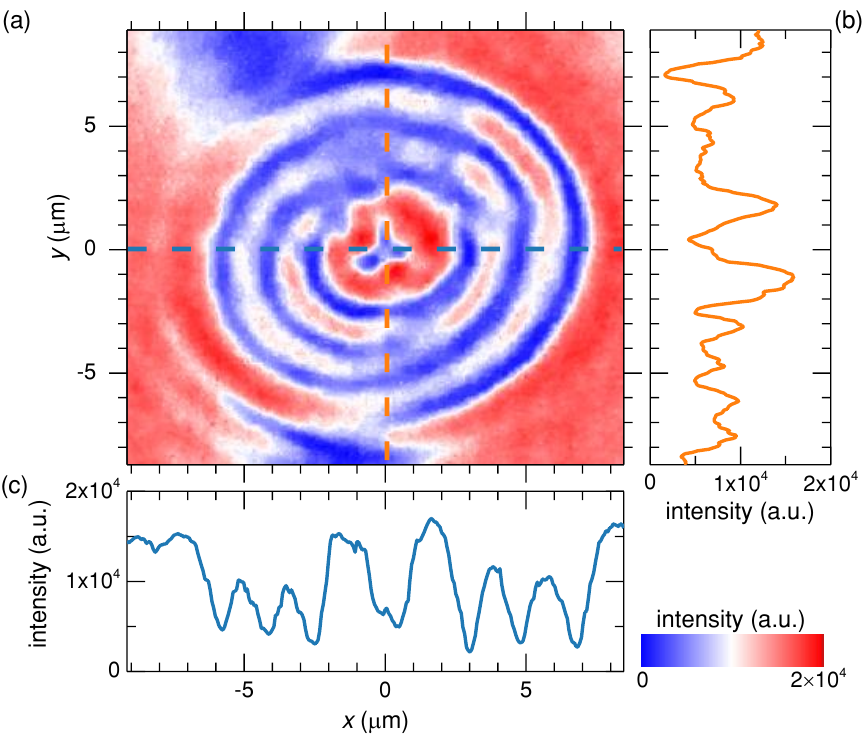}
	\caption{(a)\,Measured interferogram. (b)\,Intensity profile obtained from the $y$-cut shown in (a) in orange dashed line. (c)\,Intensity profile obtained from the $x$-cut shown in (a) in blue dashed line.}
	\label{fig:FiberA14_Interferometer}
\end{figure}
To determine the elevation $z$ as a function of position, we use interferometry: a HeNe laser is split into two beams. One beam is sent to the fiber tip, the other is used as a reference. In \Fig{fig:FiberA14_Interferometer}(a) we show the interferogram we measured for the fiber used in the experiments, with corresponding $x$- and $y$- line profiles shown in (c) and (b), respectively. Using coordinate transformation\,\cite{GeAO01}, we calculate the phase $\varphi$, from which we extract the elevation using $z=(\lambda/4\pi)\varphi$, where we consider that the light travels back and forth before recombining with the reference. The corresponding 2D profile of the elevation is shown in \Fig{fig:FiberA14_Phase}(a). Two line cuts, along $x$ and $y$, are shown in \Fig{fig:FiberA14_Phase}(c) and (b) respectively. We fit the profile along $x$ using the fitting function $d(x)$:
\be
d(x)=d_0 - t_x{\rm e}^{-\frac{(x-x_0)^2}{\sigma_x^2}}\, ,
\label{eq:dimple_fit_function}
\ee 
with the same expression used also along $y$.
The in-plane diameter $D$ of the dimple is given by $D_{i}=2\sigma_i$, while $t_i$ described the elevation at center of the dimple, with $i=x,\, y$. $d_0$ represents a total offset, and it is very important for determining the correct profile. In our analysis, $d_0$ is determined from the brighter region at the bottom right corner of \Fig{fig:FiberA14_Phase}(a). Indeed, moving from this region towards the center, the elevation $z$ decreases gradually, without showing any abrupt jump due an incorrect phase unwrapping method. From the fit, shown in \Fig{fig:FiberA14_Phase}(b) and (c) by dashed lines, we obtain $D_x=13.36\,\mu$m, $t_x=1.07\,\mu$m, $D_y=11.92\,\mu$m and $t_y=1.11\,\mu$m. 
We consider an effective diameter $\langle D \rangle$ and elevation $\langle t \rangle$ calculated as the average between the $x$ and $y$ measurements, and estimate tha radius of curvature $R$ of the fiber as\,\cite{HungerNJOP10}
\be
R=\frac{\langle D \rangle^2  }{8\langle t \rangle}\,.
\label{eq:ROC}
\ee 
We obtain $R=18.33\pm 0.31\,\mu$m. 

\begin{figure}[]
	\includegraphics[width=0.9\columnwidth]{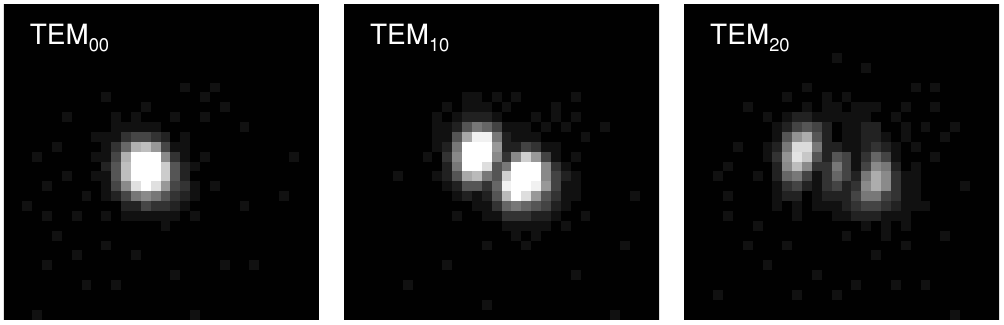}
	\caption{Imaging of the first three cavity modes measured in transmission with the laser wavelength fixed to $\lambda=833\,$nm.}
	\label{fig:CavityModeImaging}
\end{figure}
To verify formation of cavity modes with transverse confinement, we image the cavity transmission at $\lambda=833$\,nm onto a CMOS camera (Thorlabs DCC1545M), while scanning the cavity length. In \Fig{fig:CavityModeImaging} the first three transverse modes are shown, which nicely show the expected shape of Hermite-Gaussian modes. In all the experiments presented in this manuscript, we work with the fundamental TEM$_{00}$ mode.

\subsection{Sample characterization}
\subsubsection{Resonant transmission}
\label{subsec:resonant_transmission}
\begin{figure}[]
	\includegraphics[width=\columnwidth]{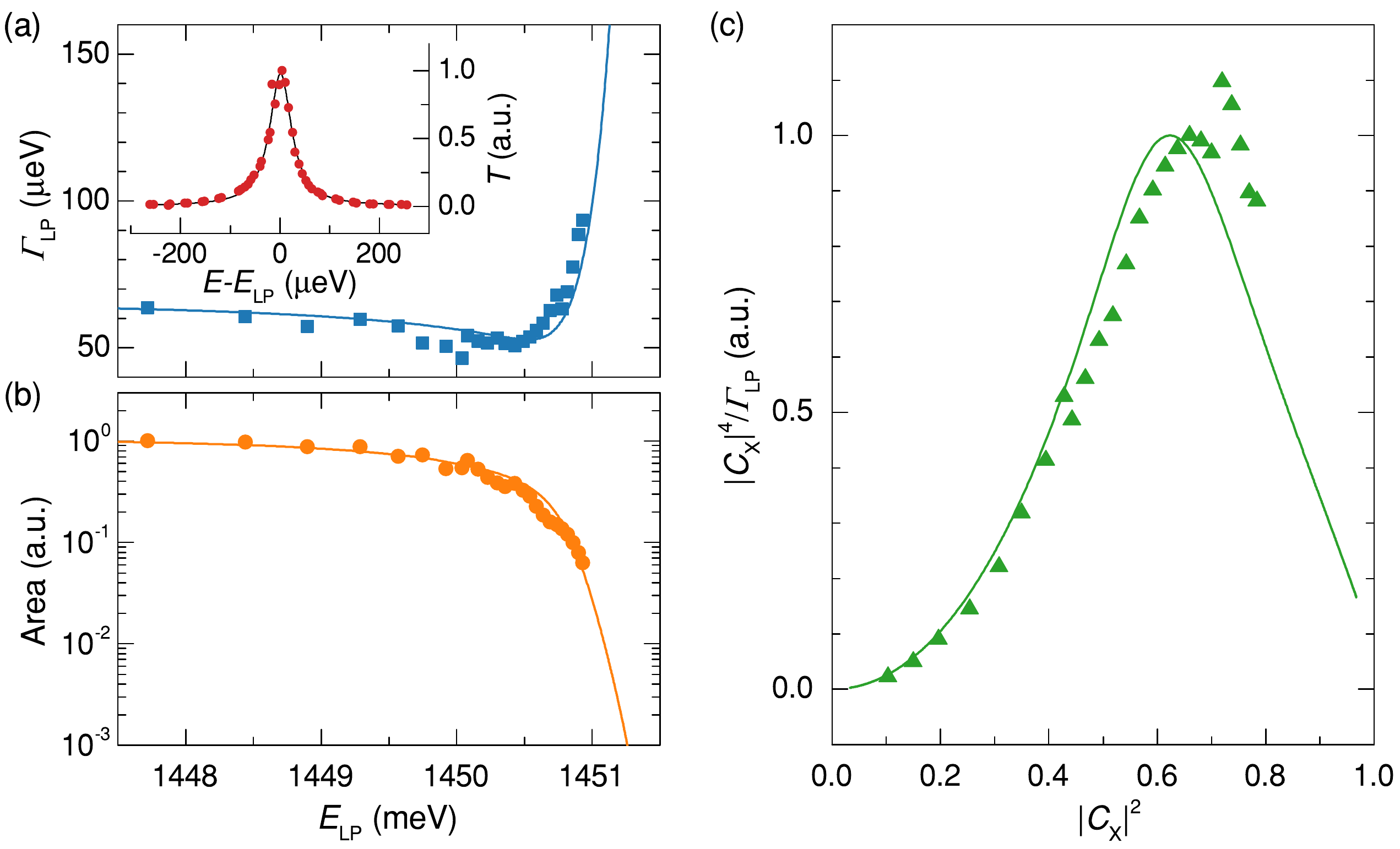}
	\caption{(a)\,,(b): LP peak linewidth and area as a function of the LP peak energy obtained from the Lorentzian fit of laser transmission spectra (an example is given in the inset), respectively. (d)\,Ratio $|\Cx|^4/\glp$, proportional to the amount of polariton-polariton interaction. Solid lines in (a),(b) and (c) are the results from the model.}
	\label{fig:characterisation}
\end{figure}

To better characterize the LP mode, we perform high resolution laser transmission scans across the peak resonance for different piezo voltages, with the transmitted photons recorded by a superconducting nanowire single photon detector (SNSPD). An example of transmission spectrum at $\Vp=45$\,V is shown in the inset of \Fig{fig:characterisation}(a), together with a Lorentzian fit. The peak linewidth $\glp$ (full-width at half-maximum (FWHM)) and area obtained from the fit are shown in \Fig{fig:characterisation}(a) and (b) respectively, as a function of the LP peak energy $\Elp$. We note that in the following we use the upper case $\glp$ to distinguish the linewidth extracted from resonant transmission from $\glpem$, which instead describes the SE linewidth. For $\Elp<1448$\,meV, corresponding to very low excitonic content, the linewidth is approximately 64\,$\mu$eV. As we increase the energy by reducing the cavity length, the linewidth slightly decreases down to $50\,\mu$eV at $\Elp=1450.5$\,meV, with the corresponding transmission dropping by about a factor of 3. For $\Elp>1450.5$\,meV, the linewidth increases rapidly,  while the transmission drops by more than one order of magnitude. Using the measured linewidth, we also determine the ratio $|\Cx|^4/\glp$, which in the limit of small nonlinearity and according to the simplest model of polariton interactions, is proportional to the interaction strength $g$\,\cite{VergerPRB06}. The results are shown in \Fig{fig:characterisation}(c). The ratio increases up to about $|\Cx|^2=0.7$, while for larger values of $|\Cx|^2$ it decreases due to the broadening of the LP linewidth. 

To describe this behaviour, we use the model proposed by Diniz et al. in \Onlinecite{DinizPRA11} for an inhomogeneously broadened ensemble of emitters strongly coupled to a cavity, described in \Sec{sec:Strong coupling simulations}, also used in \Onlinecite{DelteilNMa19} for a similar analysis. We used the experimentally determined Rabi splitting, the cavity linewidth $\kappa$ of 64\,$\mu$eV measured at low $|\Cx|^2$, and an exciton inhomogeneous broadening $\sigma=435\,\mu$eV. A homogeneous broadening coming from a damping rate $\gx=40\,\mu$eV is included to take into account additional excitonic losses. This last term was also used in \Onlinecite{DelteilNMa19}, and attributed to disorder-induced losses. In \Fig{fig:characterisation}(a)\,,(b) and (c) we show the result of the model in solid lines. The model well describes all the main features of the experimental data. Importantly, from this set of measurements, we do not observe signatures of trion-induced losses, contrary to what was observed in \Onlinecite{MatutanoNMa19} using another sample structure, which is a significant advantage for correlation measurements. We tentatively explain the absence of trion-induced losses in this new sample structures as being a result of reduced residual doping due to improved growth conditions.

\subsubsection{Spontaneous emission regime}
\label{subsec:SE}
\begin{figure}[]
	\includegraphics[width=0.9\columnwidth]{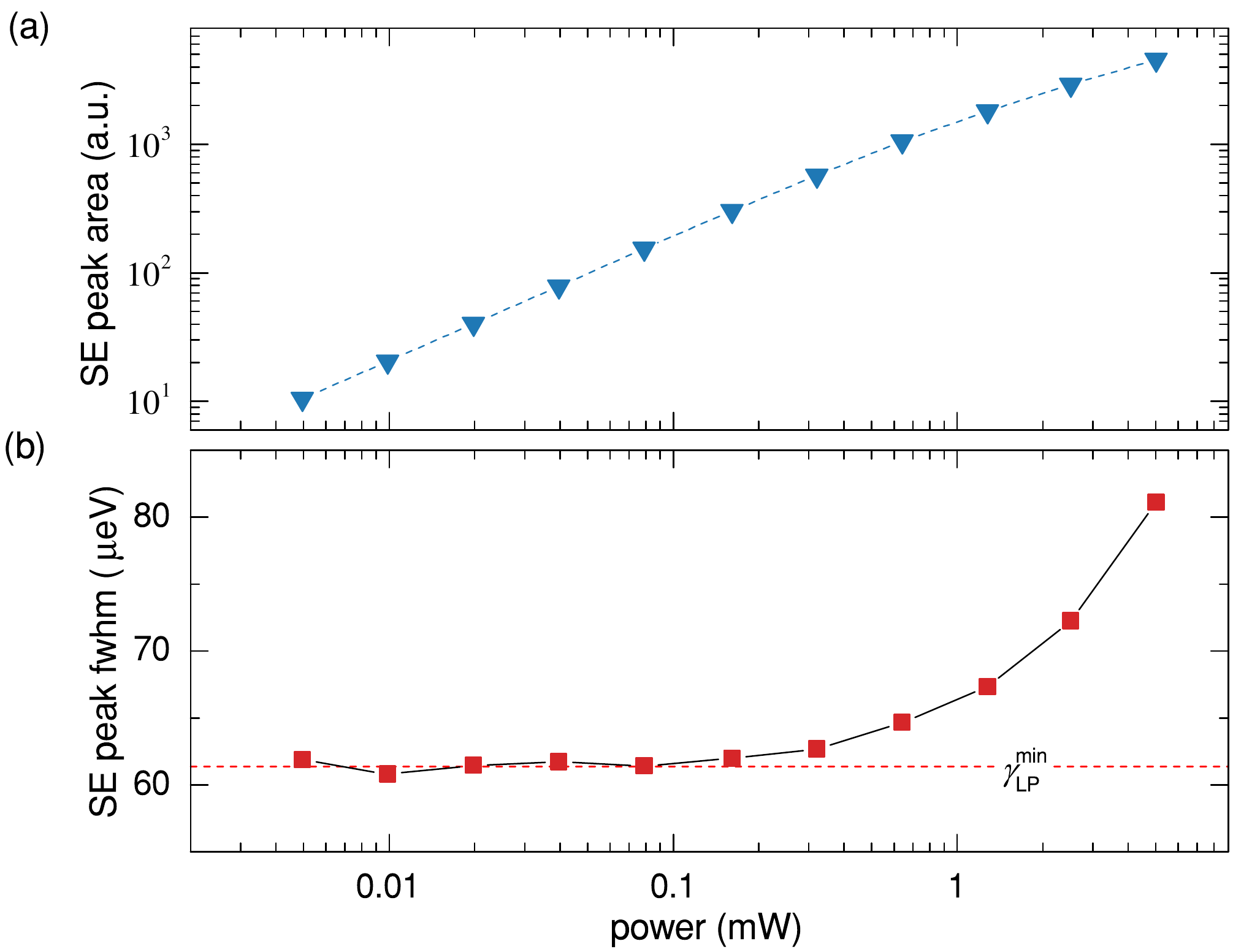}
	\caption{SE emission peak area (a) and full-width at half maximum (FWHM) (b) as a function of the excitation power. The spectra were measured at $|\Cx|^2=0.45$, corresponding to $\Elp=1450.36$\,meV.}
	\label{fig:threshold}
\end{figure}
We are interested in working in the SE regime, which is obtained at
power well below the polariton condensation threshold. In these conditions, the average polariton number is below 1, so that the intracavity polariton number states that matter the most in the experiment are $|0\rangle$, $|1\rangle$ and $|2\rangle$. Without filtering, owing to its incoherence, the corresponding emitted light generally exhibits a chaotic statistics and hence a $\gtwo(0)>1$\,\cite{KasprzakPRL08,AdiyatullinAPL15,KlaasPRL18}. To find the SE regime, we perform power-dependent measurements using a weak CW non-resonant laser fixed at 844\,nm (corresponding to about 20\,meV above the polariton level) and recording the emission with a spectrometer. We chose this excitation energy to reduce population of impurities and therefore minimising potential losses. We note, however, that the excitation energy is well within the stop-band of the distributed Bragg reflectors, so that the power reaching the sample is much lower than the nominally measured one. By fitting the LP peak with a Lorentzian, we obtain the peak area and linewidth $\glpem$, which are shown in \Fig{fig:threshold}(a) and (b), respectively, for $|\Cx|^2=0.45$. $\glpem$ remains approximately constant to a value $\glpemmin$ of about 62\,$\mu$eV, up to a power $P=1\,$mW. For $P >1$\,mW, it slowly increases. At the onset of lasing, the linewidth is expected to decrease due to the coherence build-up\,\cite{KasprzakN06}. As we do not observe this behaviour, we conclude we do not reach the lasing regime within the measured power range. This is further confirmed by the trend of the peak area, which does not show any threshold behaviour. Similar behaviour was observed also for other excitonic fractions (see \Sec{subsubsec:comparison with experiment}), although different values of $\glpemmin$ were found. For consistency, all the correlation measurements described below are acquired using an excitation power for which $\glpem \approx 1.1 \glpemmin$. Such a choice allows to minimise power broadening, which could destroy correlations, without sacrificing the photon count rate. As we will show later, in this power range we expect to work with an average polariton number $\nav \approx 0.03$. 

\subsection{Polariton anticrossing}
\label{subsec:anticrossing}
\subsubsection{Fit of the polariton anticrossing}
\label{subsec:coupled_oscillator_model}
\begin{figure}[]
	\includegraphics[width=0.9\columnwidth]{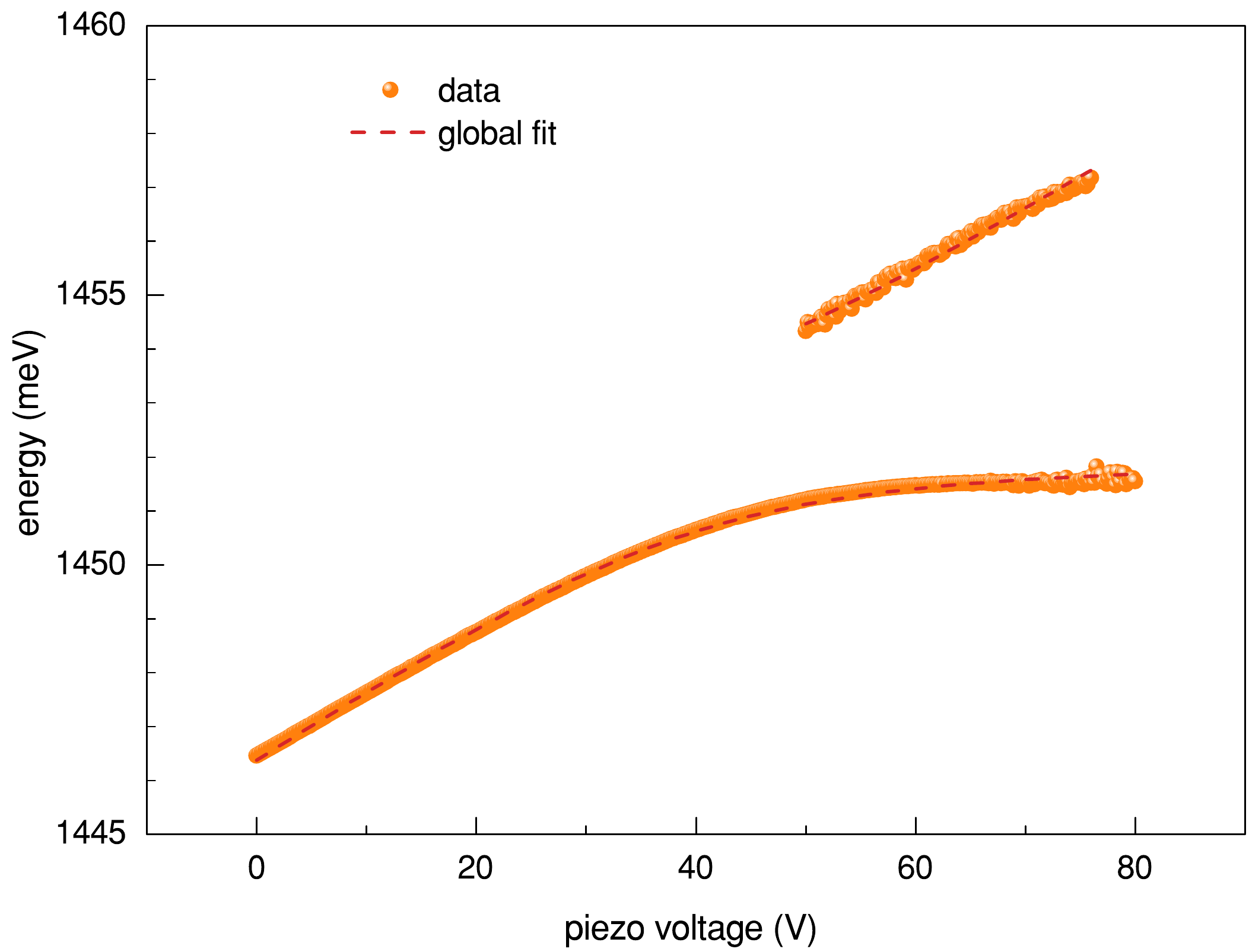}
	\caption{LP and UP peak centers plotted as a function of the piezo voltage. Dashed line is a fit to the data.}
	\label{fig:AnticrossingFit}
\end{figure}
From \Fig{fig:ExperimentalArrangement}(d), we first determined the LP and UP peak energies by fitting the spectra around the corresponding energy range using a Lorentzian function. For the UP, we fit only data for $50 < \Vp < 76\,$V, where the UP is visible. The results are shown in \Fig{fig:AnticrossingFit}. In order to fit these data, we use a standard coupled-oscillator model:
\be
E_{{\rm LP,UP}} = \frac{\left(\omega_{\rm c} + \omega_{X}\right)}{2} \pm \sqrt{\Rabi^2 + \frac{\delta}{2}^2}\, ,
\label{eq:coupled_oscillator_model}
\ee
where $\delta$ is the detuning between the bare exciton $\omega_{X}$ and bare cavity $\omega_{\rm c}$, and $\Rabi$ is the half vacuum Rabi splitting. The bare cavity energy is a function of the cavity length $L$, and for the fundamental ${\rm TEM}_{00}$ mode is given by
\be
\omega_{\rm c} = \frac{c}{2L}\left[ 2\pi q + {\rm acos}\left(\sqrt{1-\frac{L}{R}}\right) + \phi\right]
\label{eq:bare_cavity}
\ee
with $c$ the speed of light, $q$ the integer longitudinal mode quantum number and $\phi$ and additional phase term taking into account the penetration depth into both DBRs. Finally, we need to relate the cavity length to the applied voltage $\Vp$. We found a better fit by considering a non-linear dependence as in the following:
\be
L = L_0 - s_{1}\Vp - s_{2}\Vp^2\, ,
\label{eq:length_voltage_dependence}
\ee
with $L_0$ the cavity length at zero voltage and $s_1$ and $s_2$ the linear and non-linear coefficients respectively. With all these ingredients, we perform a global fit to the anticrossing data, and the result is shown in \Fig{fig:AnticrossingFit} in dashed line. From the fit, we obtain: $L_0=20.308\pm 0.007\,\mu$m, $R=21.0\pm 0.2\,\mu$m, $\hbar \omega_X=1452.08\pm 0.1$\,meV, $\Omega=1.52\pm 0.45$\,meV, $\phi=6.25\pm 0.03$\,rad, $q=6.319\pm 0.003$, $s_1=2.09\pm 0.05 \times 10^{-3}\,\mu {\rm m}/{\rm V}$ and $s_2=(-1.3\pm 0.6)\times 10^{-6}\,\mu{\rm m}/{\rm V}^2$. We note that the phase $\phi$ is proportional to the penetration depth of the cavity mode field into both DBRs. If, for example, we have a total penetration depth $L_{\rm p}$, the corresponding phase would be $2\pi L_{\rm p}/\lambda$. The estimated total penetration, using nominal expressions\,\cite{BesgaPRAp15}, and a central wavelength $\lambda=0.855\,\mu$m, is $0.71\,\mu$m. Using the phase $\phi$ obtained from our fit, we obtain a total penetration depth of about $0.851 \pm 0.004\,\mu$m, which is quite close to the nominal value. On the other hand, there is a small discrepancy between the radius of curvature $R$ obtained from the LP and UP fit and the one obtained from the analysis presented in \Sec{sec:Fiber_Mirror}. The reason could be the non-linear behaviour of the cavity length, which might introduce some systematic error in the fit. 

\subsubsection{Strong coupling simulations} \label{sec:Strong coupling simulations}
\begin{figure}[]
	\includegraphics[width=0.9\columnwidth]{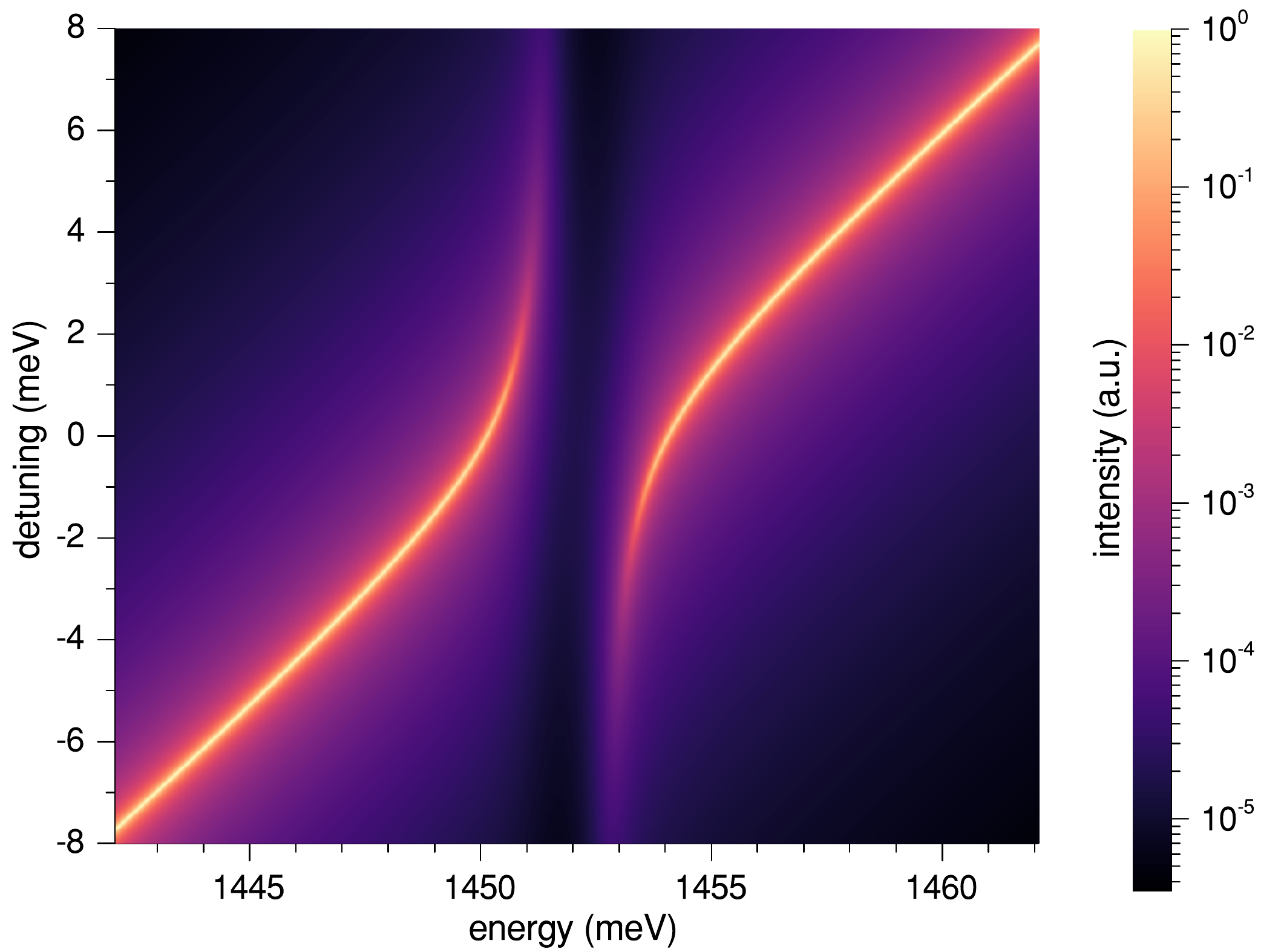}
	\caption{Cavity transmission simulated with \Eq{eq:DinizCavityTransmission}, using $\hbar\kappa=64\,\mu$eV, $\hbar\sigma=435\,\mu$eV, $\hbar \gamma_0=40\,\mu$eV, $\hbar\Omega=1.52\,$meV and $\hbar \oex=1452.08$\,meV.}
	\label{fig:anticrossing_theory}
\end{figure}
To model the data shown in \Fig{fig:characterisation}(b), (c) and (d) in the main text, we use the model presented by Diniz et al. in \Onlinecite{DinizPRA11} for an inhomogeneously broadened ensemble of emitters strongly coupled to a cavity. According to this model, the cavity transmission $t(\omega)$ is given by
\be
t(\omega) = \frac{\kappa/2i}{\omega-\omega_0+i\kappa/2 -W(\omega)}\, ,
\label{eq:DinizCavityTransmission}
\ee
with
\be
W(\omega)=\Omega^2 \int_{-\infty}^{+\infty}\frac{\rho (\omega') { d}\,\omega'}{\omega - \omega' + i\gx/2}\, ,
\label{eq:spectral_density}
\ee
$\kappa$ the cavity losses, $\gx$ the excitonic losses and $\rho$ a spectral distribution function describing exciton inhomogeneous broadening. Assuming a Gaussian distribution of transition frequencies $\rho(\omega) = \frac{\sqrt{({\rm ln}2)}}{\sigma \sqrt{\pi}}e^{-\omega^2{\rm ln}2/\sigma^2}$, with $\sigma$ the half-with at half-maximum, $W(\omega)$ takes the form
\be
W(\omega) = -i\frac{\sqrt{{\rm ln}2}\Omega^2}{\sigma}\sqrt{\pi}e^{-\left(\frac{\omega + i\gx/2}{\sigma/\sqrt{{\rm ln}2}}\right)^2}{\rm erfc}\left(-i\frac{\omega + i\gx/2}{\sigma/\sqrt{{\rm ln}2}}\right)\, ,
\label{eq:Gaussian_broadening}
\ee
where erfc is the complex error function. In \Fig{fig:anticrossing_theory} we show the cavity transmission of \Eq{eq:DinizCavityTransmission} as a function of energy $E$ and for different exciton-cavity detunings (simulation parameters are given in figure caption). For each detuning, we fit the LP lineshape with a Lorentzian function, to extract the peak area, linewidth and the ratio $|\Cx|^4/\glp$ shown in \Fig{fig:characterisation}(b)\, ,(c) and (d), respectively. 

\subsection{Polarization splitting of the cavity mode}
\begin{figure}[]
	\includegraphics[width=0.9\columnwidth]{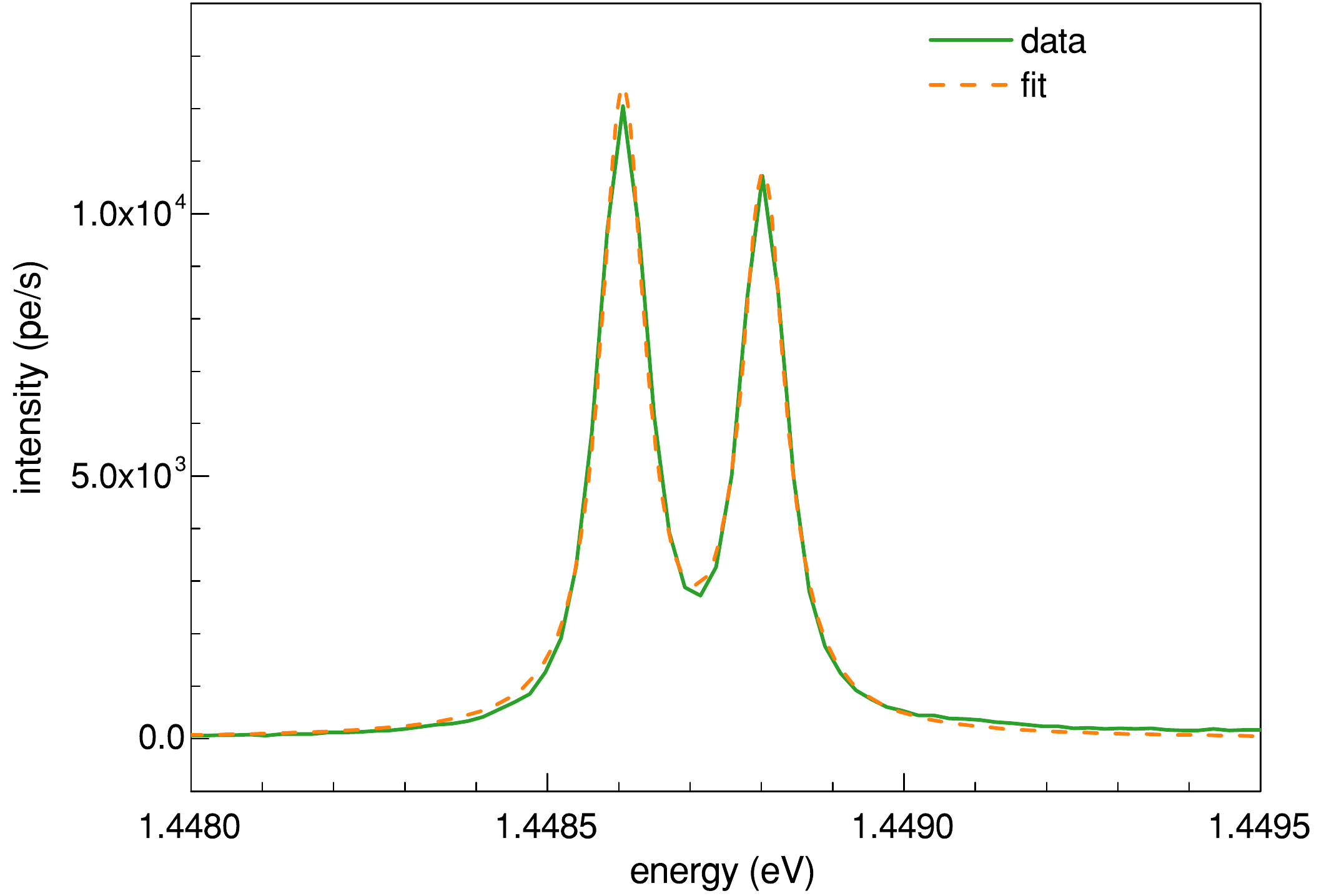}
	\caption{Polarization splitting of the fundamental TEM$_{00}$ mode measured in white light transmission. Solid line: data; dashed line: fit.}
	\label{fig:ModePolarisationSplitting}
\end{figure}
Due to residual birefringence of the GaAs, the fundamental cavity mode splits into two cross linearly polarized components. In \Fig{fig:ModePolarisationSplitting} we show the cavity transmission spectrum, zoomed around the LP emission, measured at very low exciton Hopfield coefficient. We perform a global fit using two Lorentzian peaks with the same linewidth and from the extracted peak centers, we obtain a polarization mode splitting of $190.0\pm 0.4\,\mu$eV.

\subsection{Detector response time} 
\label{sec:detector_response}
\begin{figure}[]
	\includegraphics[width=0.9\columnwidth]{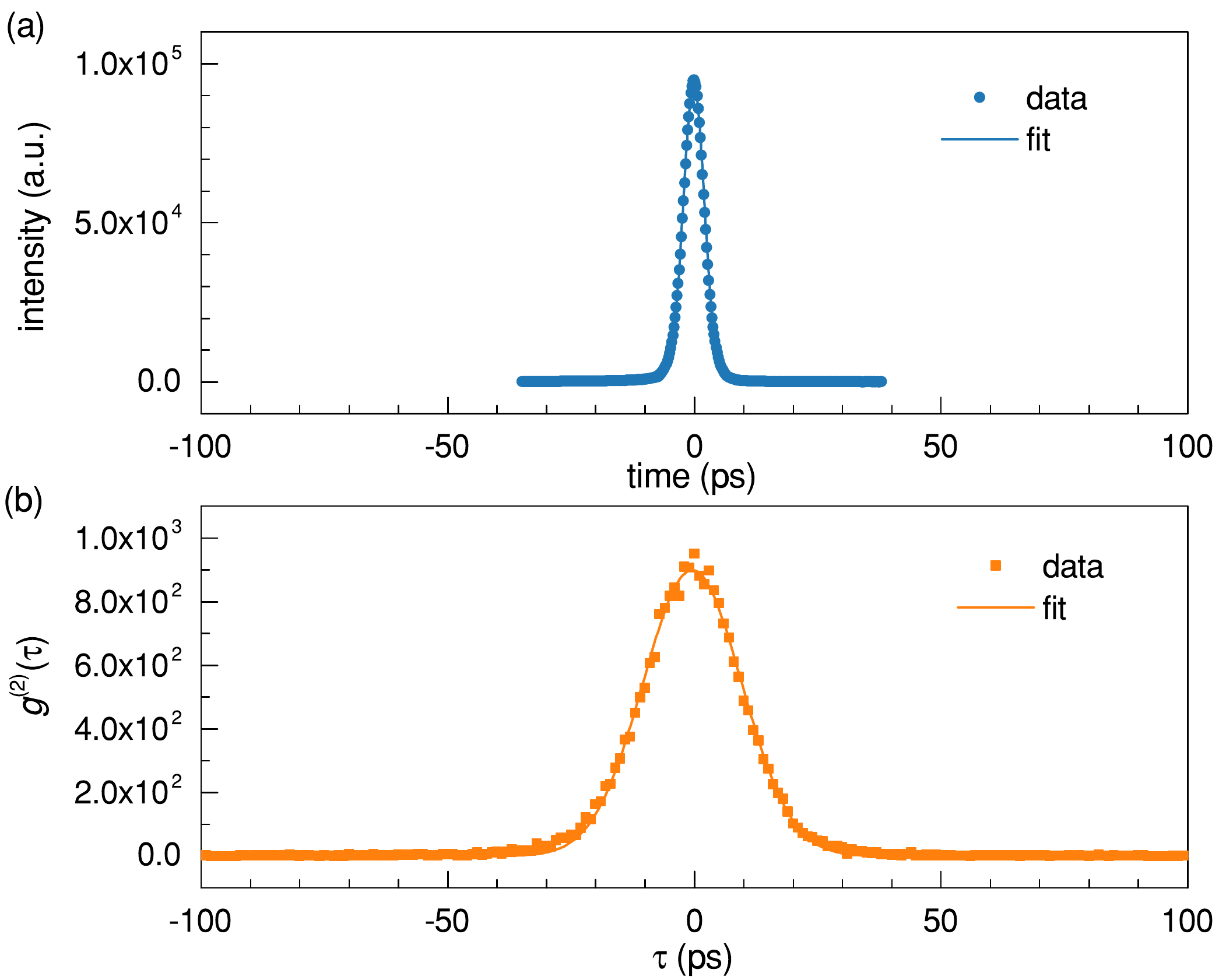}
	\caption{(a)\,Laser pulse duration measured with the streak camera. (b)\,$\gtwo(\tau)$ of the laser pulse measured with SSPDs detectors connected to the time tagger.}
	\label{fig:DetectorResponse}
\end{figure}
The knowledge of the time response of our HBT detection setup is very important for a quantitative comparison between experiment and theory. In our experiments, the response time is determined by the time jitter of the SSPDs detector and the time tagger used for counting the electrical pulses generated by the detectors upon arrival of a photon. For time tagging, we use the Swabian Instruments HiRes model. To measure the total system response, we measure the $\gtwo$ of laser pulses centered at 840\,nm. The time duration of the pulses is determined with a separate measurement using a streak camera. The time profile of the pulses is shown in \Fig{fig:DetectorResponse}(a). From a Gaussian fit we obtain a duration of $\sigma_{\rm pulse}=5.01\pm 0.01$\,ps (FWHM). The corresponding $\gtwo(\tau)$ is shown if \Fig{fig:DetectorResponse}(b), and using a Gaussian fit, we determine a total duration $\sigma_{\rm \gtwo}=23.54\pm 0.09$\,ps. As the $\gtwo(\tau)$ profile is given by the convolution of the laser pulse duration and the total detector response $\sigma_{\rm det}$, $\sigma_{\rm \gtwo}^2=\sigma_{\rm det}^2 + \sigma_{\rm pulse}^2$, from which we determine $\sigma_{\rm det}=22.97\pm 0.01$ FWHM.

\subsection{Spectral filter}
\begin{figure}[]
	\includegraphics[width=0.9\columnwidth]{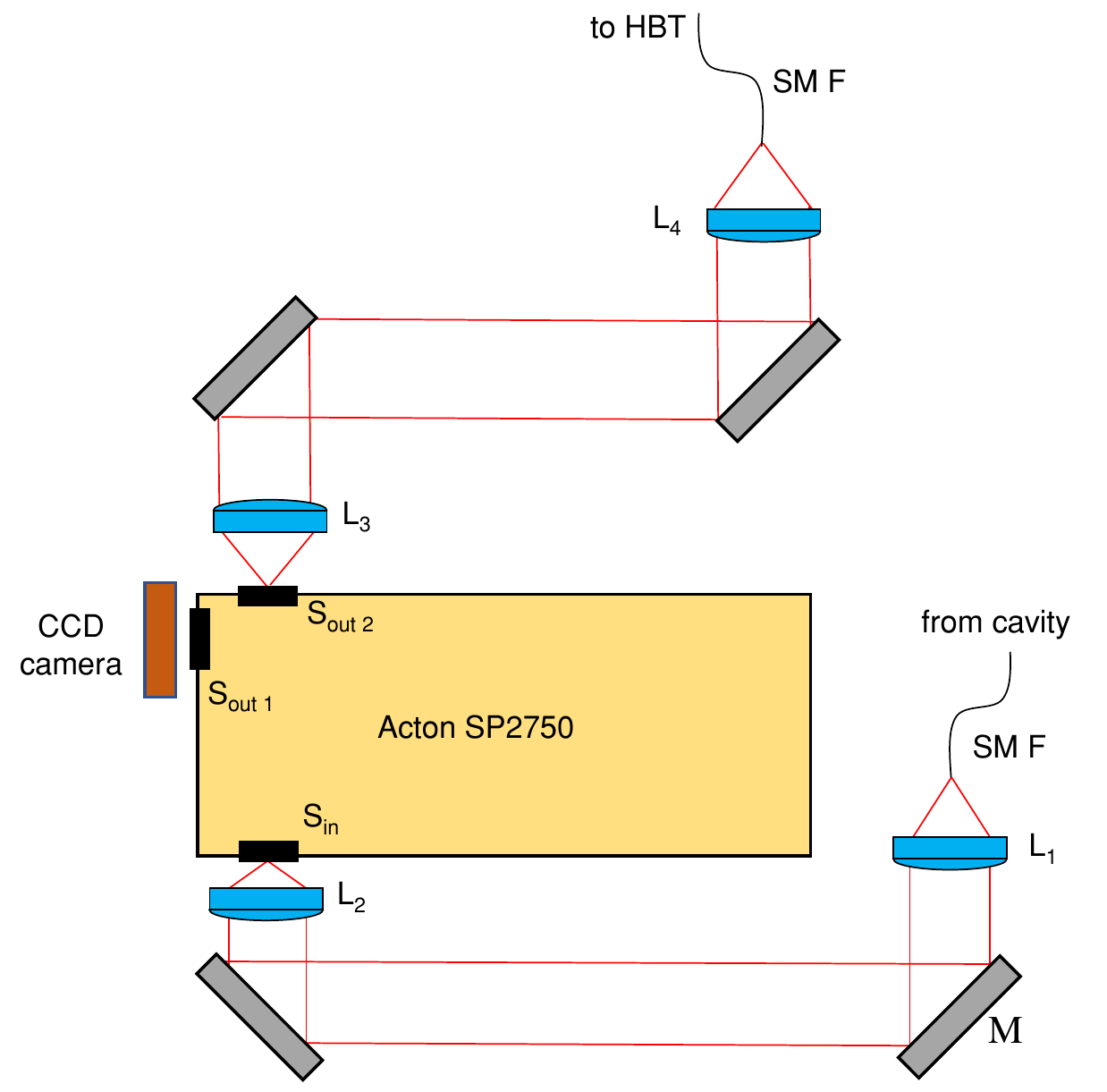}
	\caption{Sketch of the spectral filter used in the experiment. M: mirror; L: lens, S: slit; SM F: single-mode fiber.}
	\label{fig:FilterSketch}
\end{figure}
\begin{figure}[]
	\includegraphics[width=0.9\columnwidth]{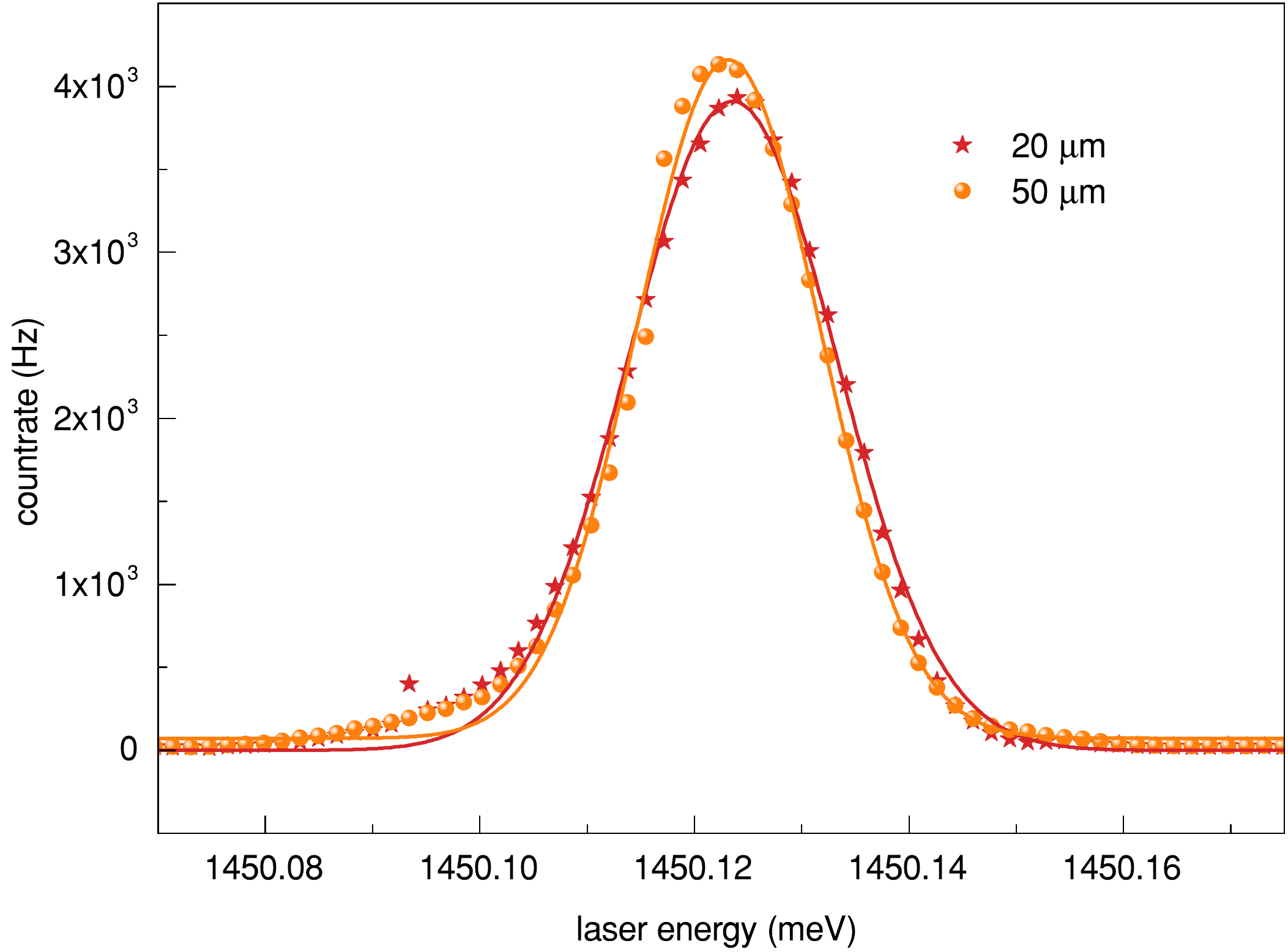}
	\caption{Line shape of the spectral filter for two different slit sizes.}
	\label{fig:SpectralFilter}
\end{figure}
A detailed sketch of the spectral filter is given in \Fig{fig:FilterSketch}. The light from the cavity is delivered to the filter via a single-mode optical fibre (SM F), SM800 from Thorlabs, which has a mode diameter of about 5\,$\mu$m at 830\,nm and a numerical aperture (NA) of about 0.14. The output light is collimated using the lens $L_1$, with a focal length $f_1=$11\,mm, resulting in a beam diameter $D=2{\rm NA}f=$3\,mm. The light is then focused onto the spectrometer input slit (S$_{\rm in}$) using a lens (L$_2$) with $f_2$=30\,mm. The f-number is therefore $f/\#=30/3=10$, which is very close to the nominal one of the spectrometer $f/\#=9.7$. This allows us to work at the maximum obtainable resolution, without losing transmission efficiency. Using a motorised removable mirror within the spectrometer, we can either send the light to a CCD camera via the output slit S$_{\rm out 1}$, or to another SM800 single mode optical fiber via a second output slit S$_{\rm out 2}$. In the latter case, the light after S$_{\rm out 2}$ is collimated by a lens (L$_3$) with $f_3=75$\,mm and placed at one focal length from the output slit, and then focused onto the fibre by another lens (L$_4$) with $f_4=11$\,mm, resulting in a high-resolution monochromator (i.e. our spectral filter). In a situation where the spectral resolution of the spectrometer is limited by the diffraction pattern of the grating, the spectral resolution should not change when changing the input slit size. To verify this is actually the case for our experimental arrangement, we measure the transmission of a laser through the filter as a function of the laser photon energy for different slit sizes, with the transmitted photons recorded with the SSPDs. In \Fig{fig:SpectralFilter} we show the transmission for the slit sizes 20\,$\mu$m and 50\,$\mu$m. As expected, we do not observe any significant variation of the filter linewidth. By fitting the lineshape with a Gaussian function (we note that this is an approximation, as the lineshape should be determined by the diffraction pattern of the largest aperture\,\cite{DemtroderBook08}), we can estimate the filter linewidth to be about 23\,$\mu$eV FWHM.

\subsection{Calculation of $\gtwo(0)$ shown in the main text}
\label{sec:g2_calculation_main_text}
\begin{figure}[]
	\includegraphics[width=0.9\columnwidth]{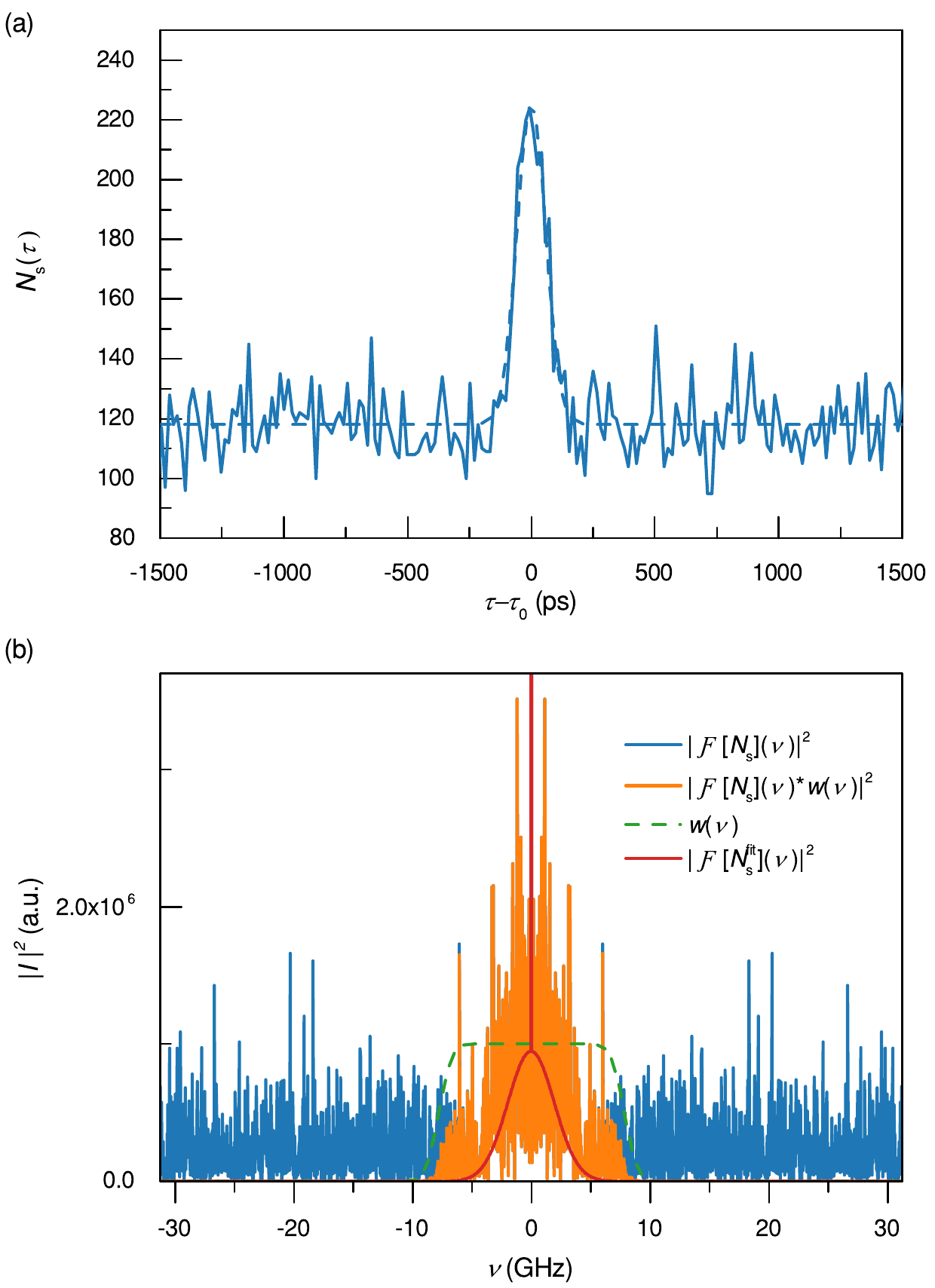}
	\caption{(a)\,Sum of raw data coincidence counts $\Ns(\tau)$ versus time delay. Dashed line is a Gaussian fit. (b)\,Blue solid line: FT of $\Ns(\tau)$; orange solid line: FT of $\Ns(\tau)$ multiplied by the window function; dashed line: window function; solid red line: FT of the Gaussian fit of $\Ns(\tau)$}
	\label{fig:g2_sum}
\end{figure}
In this section we describe the analysis we used to determine the $\gtwo(0)$ values shown in the main text. Given that all the raw data shown in \Fig{fig:Results}(a),\,(c) in the main text should have approximately the same Gaussian shape determined by the convolution of the Gaussian time response of the filter and the Gaussian time response of the detector, for each $|\Cx|^2$ we sum all the raw data to obtain a better signal-to-noise ratio, and we fitted the resulting coincidence counts $\Ns(\tau)$ with a Gaussian function. $\Ns(\tau)$ and the corresponding Gaussian fit $\Nsfit$ are shown in \Fig{fig:g2_sum}(a). From this fit, we calibrate the zero delay, so that we can keep it as a fixed parameter for the following fits. We also obtain the width $\sigma=57.64\pm 3.22$\,ps (standard deviation), which is mainly given by the filter time response convolved with the detector response. Next, we calculate the Fourier transforms (FTs) $\mathcal{F}[\Ns](\nu)$ and $\mathcal{F}[\Nsfit](\nu)$, where here $\mathcal{F}$ indicates the FT operation and $\nu$ the frequency. The results are shown in \Fig{fig:g2_sum}(b) in blue and red solid lines, respectively. $\mathcal{F}[\Nsfit](\nu)$ is also Gaussian, and this allows us to define a window function which keeps $99\%$ of the frequencies in $\mathcal{F}[\Nsfit](\nu)$, therefore rejecting high frequency noise generated by the instrument. The used window function $w(\nu)$ is shown in \Fig{fig:g2_sum}(b) in dashed line, and it is generated as the multiplication of two, frequency shifted, error functions, both of width 1.25\,GHz. Given a raw coincidence counts data $N_{\rm raw}(\tau)$, the corresponding noise filtered data $N_{\rm filtered}(\tau)$ is calculated as $N_{\rm filtered}(\tau)=\mathcal{F}^{-1}[w(\nu)\mathcal{F}[N_{\rm raw}(\tau)](\nu)](\tau)$. We fit $N_{\rm filtered}(\tau)$ using a Gaussian function with fixed centre position as per the zero delay found above. From this fit, we determine the peak amplitude $N_0$ and the background coincidence counts $Y_0$ (i.e. $N_{\rm filtered}(\tau)$ for $\tau \gg 0$) and calculate the $\gtwo(0)$ as 1 + $N_0/Y_0$. The final error on $\gtwo(0)$ is calculated using the error on $N_0$ and $Y_0$ extracted from the fit and propagated using standard error propagation. This error fixes the error bars shown in the main text. The value of $\gtwo(0)$ calculated in this way needs to be deconvolved from the detector time response, which we found to be $\sigma_{\rm det}=9.75\pm 0.04$ standard deviation (see \Sec{sec:detector_response}). The standard deviation of the deconvolved data is approximately given by $\sigma_{\rm dec} = \sqrt{\sigma^2 - \sigma_{\rm det}^2}=56.81 \pm 3.22$\,ps. This value is very similar to $\sigma$, so that the deconvolution leads to negligible relative changes, $<1\%$, in the value of $\gtwo(0)$. 

\subsection{$\gtwo(0)$ convergence plot}
\begin{figure*}[]
	\includegraphics[width=0.67\textwidth]{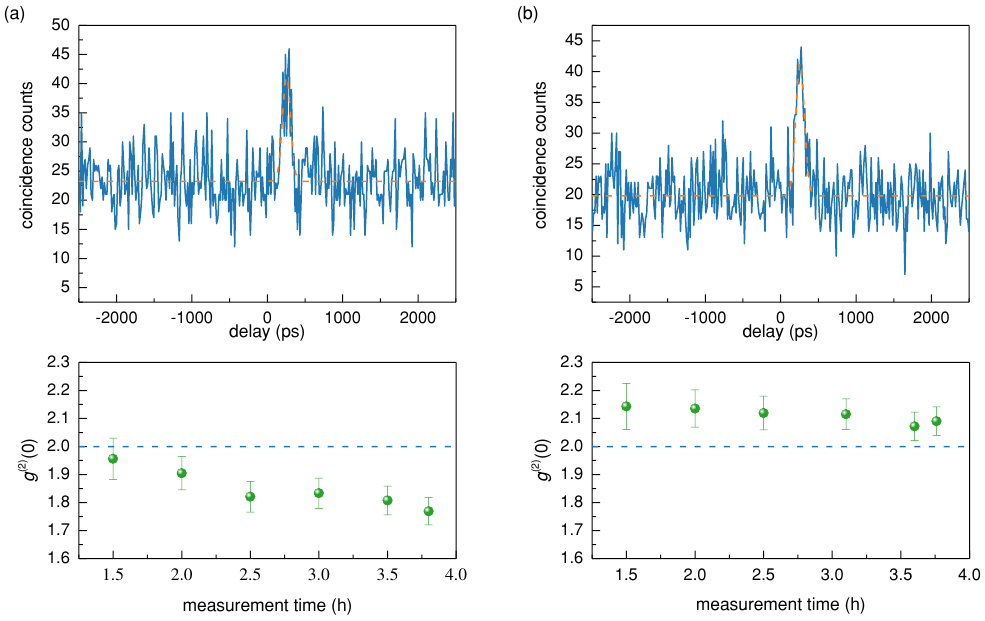}
	\caption{(a)\,Top panel: final raw coincidence counts, together with the Gaussian fit as described in \Sec{sec:g2_calculation_main_text} (dashed line), for $|\Cx|^2 = 0.42$ and $\Df=-0.6\glp$. Bottom panel: corresponding convergence plot. (b)\,As (a), but for $\Df=+0.6\glp$.}
	\label{fig:convergence_plot}
\end{figure*}
While the error bars on the $\gtwo(0)$ values extracted from the analysis described in \Sec{sec:g2_calculation_main_text} give an indication of the precision of the data, they do not say anything about their corresponding accuracy, i.e., how close they are to the true value. In order to make sure to acquire reliable and consistent measurements, for each data point, we plot the value of the $\gtwo(0)$ as a function of the acquisition time. Two examples, for $\Df=-0.6\glp$ and $\Df=+0.6\glp$ are shown in the bottom panel of \Fig{fig:convergence_plot}(a) and (b) respectively, with the corresponding final raw data shown in the top panel. From the trend, we can be sure the measurement has converged and move to the next data point. Usually, we found that convergence is obtained when a background level of about 20 coincidence counts is reached.   
\subsection{Theoretical analysis}
We use a master equation approach, with the density matrix $\density$ describing the intracavity polariton state, and $\nr$ the reservoir excitons number. We include two Linblad terms: one describing the polariton losses by leakage through the cavity mirrors
\be
\linb\density=\gsim (\annh \density \creat - \frac{1}{2}\creat \annh \density - \frac{1}{2}\density \creat \annh)\, ,
\label{eq:linbladian_loss}
\ee
and another one that describes the effective polariton pump
\be
\linbp\density=\gr \nr (\creat \density \annh - \frac{1}{2} \annh\creat \density - \frac{1}{2}\density \annh\creat )\, ,
\label{eq:linbladian_pump}
\ee
where $\nr$ is the time-dependent population of the excitonic reservoir. Taking into account the free dynamics of the LP mode, captured by Hamiltonian
\be
\hat{\cal H} = \omega_\text{LP}\creat\annh+g\creats\annhs+\gprime\creatt\annht,
\ee
the total dynamics of the mode thus reads 
\be
\dot{\density} = -\frac{i}{\hbar} \left[\Ham,\density \right] + \linb + \linbp
\label{eq:master_equation} 
\ee
and it is coupled to the reservoir dynamics via
\be
\dnr = F - \gr \nr \left(n+1\right) - \gd \nr
\label{eq:reservoir_dynamics}
\ee
where $F$ is the cw pumping, $n$ the time dependent polariton occupation number and $\gd$ is the decay rate to a dark reservoir. Note that a stimulated relaxation term is involved in agreement with the pump Lindbladian. Physically this stimulation term is a well known feature of lasers rate equations, which is equally valid and routinely used to describe polaritonic stimulated emission. In order to further explicit \Eq{eq:master_equation}, we use the fact that, owing to the non-resonant excitation mechanism, the cavity initial state can only be incoherent, i.e. it requires only the density matrix diagonal elements to be described. This remains true at any later time since neither the Hamiltonian, nor the Lindblad operators couple the diagonal and the off-diagonal elements of the density matrix. The calculation is thus confined within the density matrix diagonal subspace and \Eq{eq:master_equation} simplifies into: 
\be
\begin{split}
	\dot{\rho}_n &=\gsim \left[\left(n+1\right)\rho_{\rm n+1} - n\rho_n\right]\\
	&+ \gr \nr \left[n\rho_{\rm n-1}-\left(n+1\right)\rho_n\right]\,.\\
\end{split}
\label{eq:master_equation_explicit_SI}
\ee
with $\rho_n=\bra{n}\density \ket{n}$. \Eq{eq:master_equation_explicit_SI} is equivalent to the equation of motion of the laser radiation field density matrix from the quantum theory of the laser developed by Scully \& Lamb\,\cite{ScullyPR67}, with $\gr\nr$ describing the gain term and $\gsim$ describing the loss term, and a self-saturation coefficient equal to zero. The theory was adapted to the description of polariton systems by Wouters and Carusotto\,\cite{WoutersPRL07} in 2007, and recently revisited by Klaas et al. in \Onlinecite{KlaasPRL18}. The SE regime, far below the lasing threshold, that we use in this work is given by the condition $\gr\nr/\gsim\ll 1$. The corresponding polariton occupation probability is $\probocc=(1-\gr\nr/\gsim)(\gr\nr/\gsim)^n$, which resembles the one of a black-body cavity, for which we expect $\gtwo(0)=2$.
\begin{figure}[]
	\includegraphics[width=0.9\columnwidth]{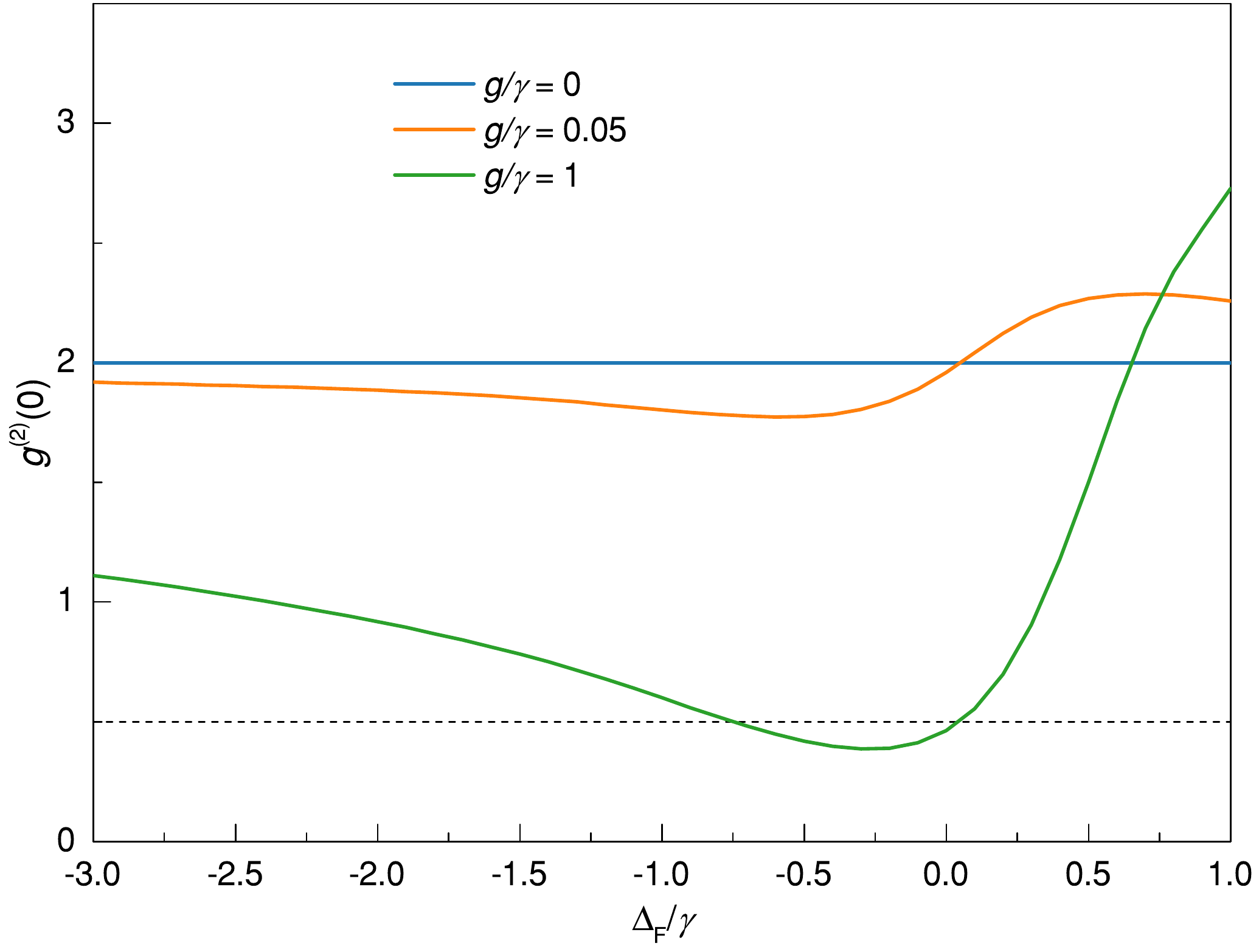}
	\caption{Influence of $g/\gamma$ on the emission statistics $\gtwo(0)$ versus filter detuning calculated using \Eq{eq:sim_g2}. Parameters: $\gsim=65.8\,\mu$eV, $\gr=1.88\,\mu$eV, $\gF=0.33\gsim$, and $\gr\nrav/\gsim=0.029$.}
	\label{fig:g2_theory}
\end{figure}
In order to compute an experimental observable such as the photon coincidence rate, we need to express the dynamics of the cavity conditioned on the measurement record of the detector, i.e. the sequence of the times at which a photon is detected. Such conditioned dynamics is obtained by ``unravelling" the master equation. In each single realization of the experiment, the system follows a different quantum trajectory, characterized by the stochastic times at which photons are detected. In order to identify the unravelling corresponding to the detection scheme we are interested in, we rewrite \Eq{eq:master_equation_explicit_SI} under the following Kraus representation:
\be
\begin{split}
	&\density(t+{d}t)=\Knj \density \Knjd +\\
	&\sum_{n}\left\{\Kndet \density \Kndetd + \Knndet \density \Knndetd + \Knp \density \Knpd \right\}\,,
\end{split}
\label{eq:Krauss_decomposition}
\ee
in which each term corresponds to one of the possible events that can take place within ${\rm d}t$ and that are relevant to the detection scheme (${\rm d}t$ is a time interval much smaller than any of the dynamics characteristic time). There are four of them and hence four Kraus operators, namely: (i) $\Kndet$, which corresponds to a transition from state $\ket{n}$ to state $\ket{n-1}$, i.e. the emission of a photon given the cavity contains $n$ polaritons, followed by a detection event (a photodetector click). (ii) $\Knndet$, which corresponds to a transition from state $\ket{n}$ to state $\ket{n-1}$, i.e. the emission of a photon given the cavity contains $n$ polaritons, which is not detected. The photon has been either rejected by the filter, or by the finite quantum efficiency of the detector. (iii) The ``no jump" operator $\Knj$ which describes the case where no photon is emitted during $dt$. And finally, (iv) the pump operator $\Knp$, which describes the addition of a polariton in the cavity by relaxation from the excitonic reservoir, given it contained $n$ polaritons at the beginning of the time step. These operators are normalized such that, for the Kraus operator $K$, the expectation value of $K^{\dagger}K$ equals the probability $P$ for the event associated with Kraus operator $K$ to occur between times, i.e. $P=\langle K^{\dagger} K \rangle$. From these considerations, we can express three out of four Kraus operators:
\begin{subequations}
	\begin{align}
	&\Kndet = \sqrt{\gsim {d}t n \eta P_{n}^{\rm F}}\ket{n-1}\bra{n}\\
	&\Knndet = \sqrt{\gsim {d}t n \left[1-\eta P_{n}^{\rm F}\right]} \ket{n-1}\bra{n}\\
	&\Knp = \sqrt{\gr \nr {d}t \left(n+1 \right)}\ket{n+1}\bra{n}\, ,
	\end{align}
	\label{eq:known_Kraus_operators}
\end{subequations}
where $\eta$ is the detector quantum efficiency and $P_{n}^{\rm F}$ the probability of the photon emitted from the state $n$ to pass through the spectral filter. This quantity is the new ingredient of our theory and it is calculated in the following way: (i) we assume that for a given number of excitations $n$, the probability that the cavity emits a photon at the frequency $\omega$ is given by $L^{\gsim}_{\omega_n}(\omega){\rm d}\omega$, where $L^{\gsim}_{\omega_n}$ is a normalized Lorentzian function of linewidth $\gsim$ and central frequency $\omega_n$. (ii) The photon then has to pass through the spectral filter, which is characterized by a transmission probability $T(\omega)=G_{\freqF}^{\gF}(\omega)$, where $G_{\freqF}^{\gF}$ is a Gaussian probability density function with center frequency $\freqF$ and linewdith $\gF$. These parameters have been determined experimentally by spectral characterization of the filter. The probability for a photon to be transmitted through the filter thus reads:
\be
P_{n}^{\rm F} = \int_{-\infty}^{+\infty} {d}\omega\, L^{\gsim}_{\omega_n}(\omega) \times G_{\freqF}^{\gF}(\omega)\,.
\label{eq:photon_transmission}
\ee
For each $n$, $P_{n}^{\rm F}$ is therefore a function of $g$ and $\omega_{\rm F}$. Finally, the no jump operator $\Knj$ is obtained by identifying \Eq{eq:Krauss_decomposition} with \Eq{eq:master_equation}, and by using the explicit Kraus operator expressions of \Eq{eq:known_Kraus_operators}:
\be
\begin{split}
	\Knj = \density  - \frac{i}{\hbar} dt[\hat{\cal H},\density]\\- \gsim {d}t\sum_{n}\frac{1}{2}n&\left( \ket{n}\bra{n}\density + \density \ket{n}\bra{n} \right)\\
	-\gr \nr {d}t\sum_{n}\frac{1}{2}(n+1)&\left( \ket{n}\bra{n}\density + \density \ket{n}\bra{n} \right)
\end{split}
\label{eq:no-jump_operator}
\ee
With these considerations, we now use two different approaches to calculate $\gtwo(0)$: we first show the analytical theory already presented in the main text. Second, we use the quantum Montecarlo method to simulate also the dynamics of the system and obtain $\gtwo(\tau)$, from which we extract the value at zero delay. We use the Montecarlo method to confirm the results of the analytical solution. 

\subsubsection{Analytical theory}
\label{subsec:Analytical_Theory}
From the Kraus representation, the detection superoperator, which models the backaction of a detection event on the polariton field, can be calculated as:
\be
\mathcal{I}[\density]=\sum_{n}\Kndet \density \Kndetd\, ,
\label{eq:det_superop}
\ee
with the expectation value $\langle \mathcal{I} \rangle={\rm Tr} \{\mathcal{I}[\density(t)]\}$ yielding the probability of a detection event at time $t$. More explicitly:
\be
\mathcal{I}\left[\density\right] = \gsim {d}t \eta \sum_n P_{n}^{\rm F}\annh \ket{n}\bra{n}\density\ket{n}\bra{n}\creat\, ,
\label{eq:det_superoperator}
\ee
The second order correlation function can be defined as
\be
\gtwo(\tau) = \lim_{t \to \infty} \frac{\langle \mathcal{I}(t+\tau) \mathcal{I}(t) \rangle}{ \langle \mathcal{I}(t) \rangle^2}\, ,
\label{eq:g2_tau}
\ee
where explicitly $\langle \mathcal{I}(t+\tau) \mathcal{I}(t) \rangle = {\rm Tr}\{\mathcal{I}e^{\linb_\text{tot}\tau}\mathcal{I}\density (t)\}$, with $\linb_\text{tot} = \linb +\linbp$   the total Linbladian. As we are interested in the zero delay value, we set $\tau=0$ and obtain:
\begin{widetext}
	\begin{equation}
	\begin{split}
	\gtwo(0) =\lim_{t \to \infty} \frac{\langle \mathcal{I}(t)\mathcal{I}(t)\rangle}{  \langle \mathcal{I}(t) \rangle^2} &= \lim_{t \to \infty}\frac{\sum_{n,m} P_{n}^{\rm F}P_{m}^{\rm F} {\rm Tr}   \left\{\annh\ket{m}\bra{m}\annh\ket{n}\bra{n}\density(t)\ket{n} \bra{n}\creat\ket{m}\ket{m}\creat\right\}}{\left( \sum_{n} P_{n}^{\rm F}{\rm Tr}   \left\{\annh\ket{n}\bra{n}\density(t)\ket{n} \bra{n}\ket{n}\creat\right\} \right)^2}\\
	& =  \frac{\sum_{n} P_{n}^{\rm F} P_{n-1}^{\rm F} n (n-1) \probocc}{\sum_{n} \left( P_{n}^{\rm F} n \probocc \right)^2}\, ,
	\end{split}
	\label{eq:sim_g2}
	\end{equation}
\end{widetext}
where we used $\lim_{t \to \infty} \bra{n}\rho(t)\ket{n}=\probocc$.
In \Fig{fig:g2_theory} we calculated $\gtwo(0)$ as a function of filter detuning $\Delta_F$ using \Eq{eq:sim_g2}. When $g/\gsim=0$ (blue curve), the polariton emission spectrum does not depend on $n$, and therefore $\gtwo$ does not depend on the filter detuning, and its value is fixed by the polariton occupation statistics $\probocc$, which is thermal, and therefore results in $\gtwo(0)=2$. On the other hand, when $g/\gsim > 0$, the polariton spectrum depends on $n$. By tuning the filter position, we thus change the relative probability of detecting the $\ket{2}\to \ket{1}$ and $\ket{1}\to \ket{0}$ emission events from the polariton ladder. The emission statistics is thus modified and depends on the filter position in frequency. The amplitude of the statistics modulation increases with the nonlinearity figure of merit. When $g/\gamma\geq 1$, a filter position is found where $\gtwo(0) < 0.5$ , corresponding to a dominant probability of single photon emission events.

\subsubsection{Quantum Montecarlo simulations}
\label{subsec:Montecarlo_Simulations}
We can validate the analytical theory by simulating the dynamics using the quantum Montecarlo trajectory method. With the Kraus operators in \Eq{eq:known_Kraus_operators} and \Eq{eq:no-jump_operator}, we can compute the probabilities $P_i$ for the corresponding event $i$ to occur between times $t$ and $t+{\rm d}t$ as $P_i=\langle K_i^{\dagger} K_i \rangle$, with $i$=\{det, nd, p, nj\}. The time interval ${\rm d}t$ is chosen so that $P_i \ll 1$ for $i$=\{det, nd, p\}. We can then evolve the trajectory by comparing $P_i$ with a random number with values between 0 and 1. During each trajectory, we keep track of the system density matrix, as well as of photon detection events.
\begin{figure}[]
	\includegraphics[width=0.9\columnwidth]{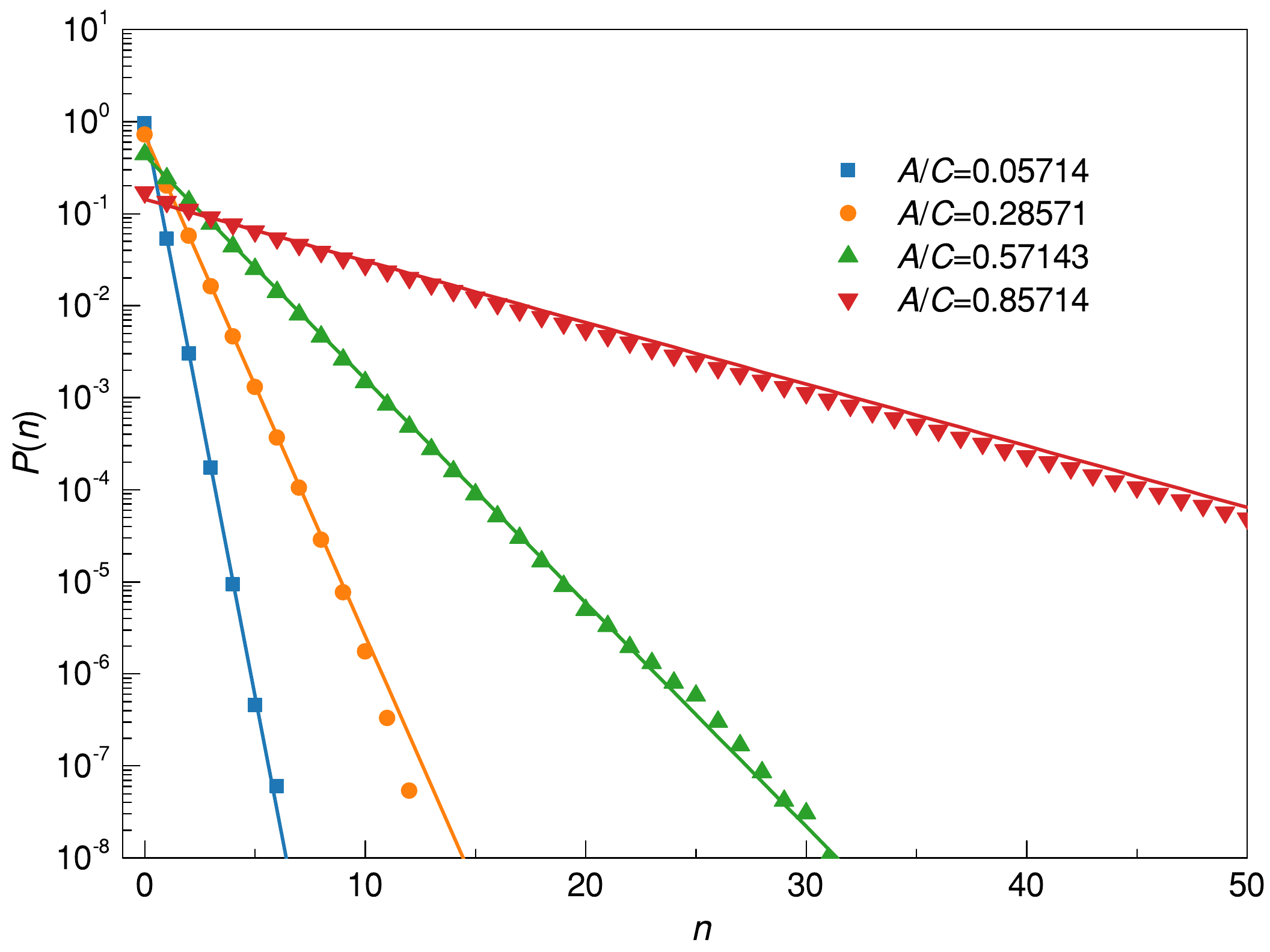}
	\caption{Symbols: polariton occupation probability $P(n)$ as a function of the polariton occupation number $n$, for different values of the threshold parameter $A/C$ below threshold. Line: model with \Eq{eq:black_body}.}
	\label{fig:Pn}
\end{figure}
\begin{figure}[]
	\includegraphics[width=0.9\columnwidth]{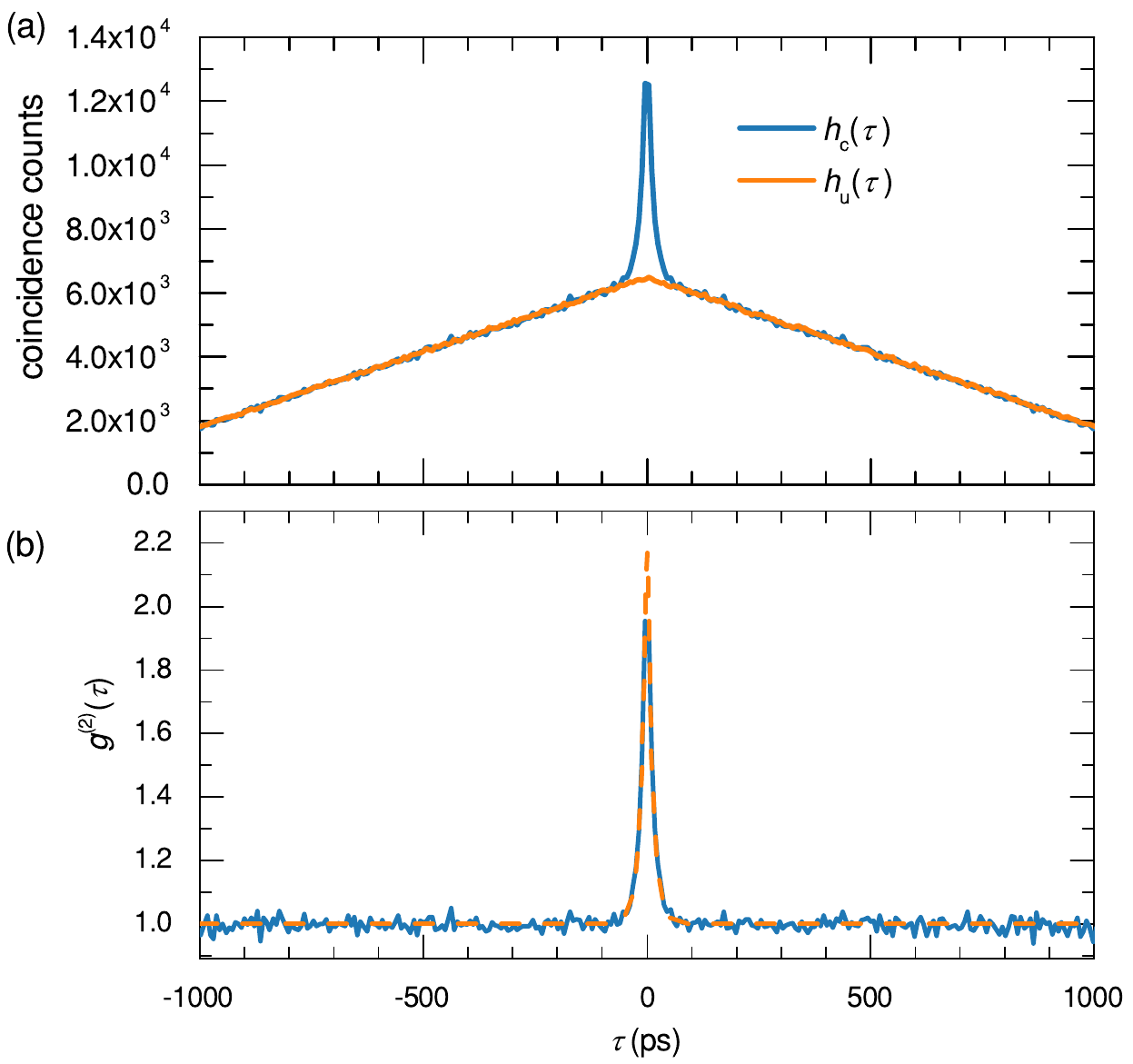}
	\caption{(a)\,Example of correlated and uncorrelated delay histograms. (b)\,Solid line: corresponding $\gtwo_{\rm sim}(\tau)$; dashed line: fit with \Eq{eq:fit_simulation_g2}.}
	\label{fig:histograms_example}
\end{figure}
\begin{figure}[]
	\includegraphics[width=0.9\columnwidth]{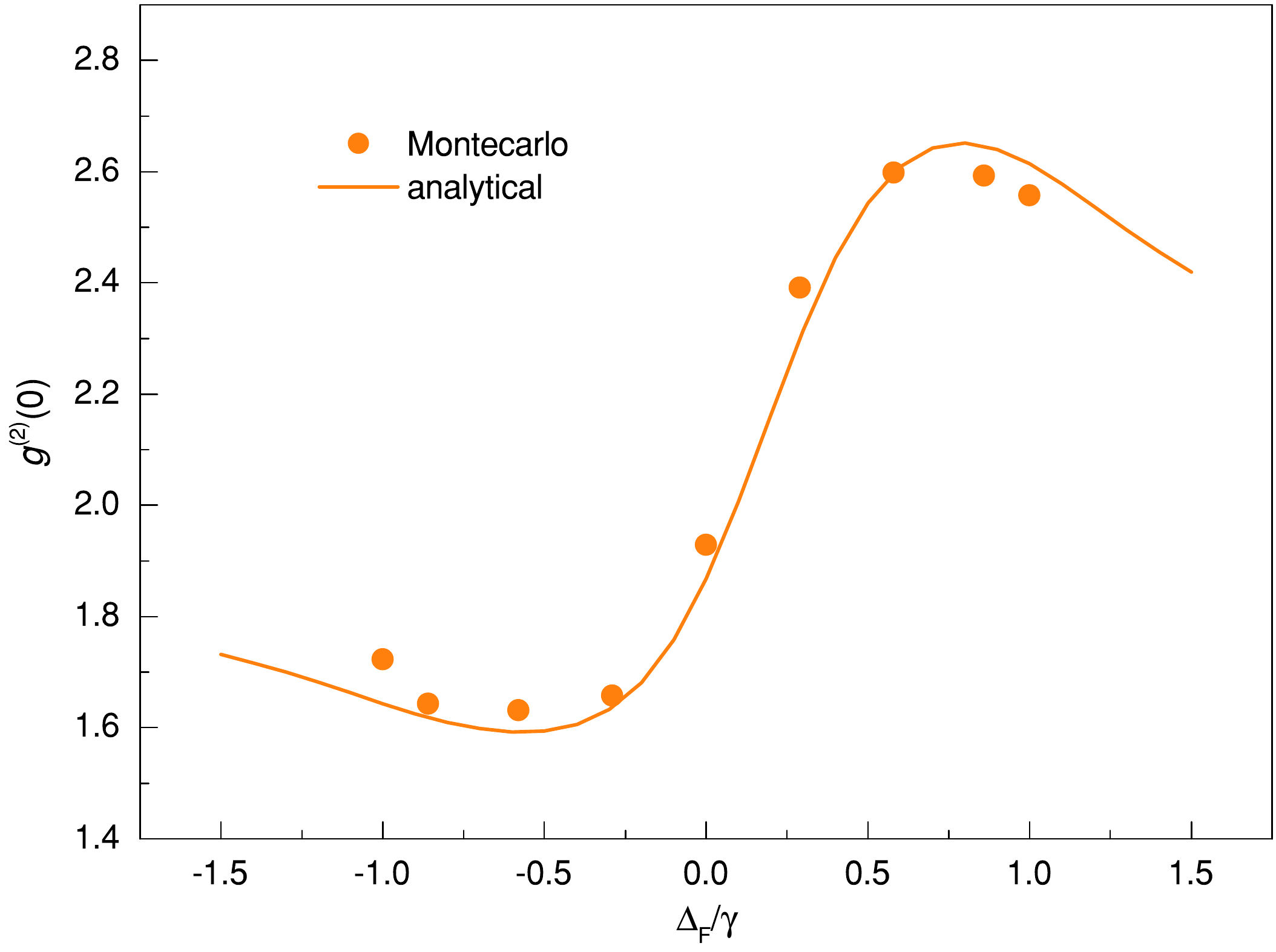}
	\caption{Comparison between Montecarlo simulations (symbols) and the analytical expression of \Eq{eq:sim_g2} for exactly the same parameters: $\nr = 5, g=0.1\,\glp$ and $\gF=0.35\glp$.}
	\label{fig:Montecarlo_vs_Analytical}
\end{figure}
\\
\noindent As mentioned above, the dynamics of the polariton field is identical to the dynamics of the photon field in the quantum theory of the laser developed by Scully \& Lamb\,\cite{ScullyPR67}. The analogy is completed by introducing a gain term $A=\gr \nr$ and a loss term $C=\gsim$. With these definitions, the polariton occupation number probability $P(n)$ below threshold ($A\ll C$) follows the statistics of a black-body cavity. In formulas:
\be
P(n)=\left[1-\frac{A}{C}\right]\left(\frac{A}{C}\right)^n\,.
\label{eq:black_body}
\ee
In \Fig{fig:Pn} we show the polariton occupation probability extracted from our Montecarlo simulations by keeping track of the polariton occupation number at each time step of the simulations. As it is possible to see, the extracted values for $P(n)$ (symbols) are well described by \Eq{eq:black_body}, as expected. \\
\noindent We calculate delay histograms using the time indices of photon detection events. We distinguish between two types of delay histograms: one for the correlated counts $\hc(\tau)$ and one for the uncorrelated counts $\hu(\tau)$, with $\tau$ the time delay. For $\hc(\tau)$ the coincidence counts are obtained using two time indices within the same trajectory, and summing over all the possible time delays. On the other hand, for $\hu(\tau)$ we use one time index from one trajectory and another time index from a second trajectory, and we then sum over all the possible combinations. Because detection events in different trajectories are not correlated, $\hu(\tau)$ provides the effective background of uncorrelated coincidence counts. Therefore, the simulated $\gtwo_{\rm sim}(\tau)$ is calculated as:
\be
\gtwo_{\rm sim}(\tau)=\frac{\hc(\tau)}{\hu(\tau)}
\label{eq:gtwo_tau_simulations}
\ee
In \Fig{fig:histograms_example}(a) we show an example of histograms with the corresponding $\gtwo_{\rm sim}(\tau)$ shown in panel (b). The characteristic triangular shape in (a) is due the fact that counts within each trajectory result in a fluctuating signal around a finite mean value, multiplied by a rectangular wave (with width equal to the total trajectory time) due to the finite time length of the simulations; when calculating the histogram counts, we are effectively convolving two rectangular pulses with the same time duration, and the result gives a triangular pulse. \\
By looking at extracted $\gtwo_{\rm sim}(\tau)$ in \Fig{fig:histograms_example}(b), we can see the time decay is exponential, rather than Gaussian as in the experiment, even though we are simulating with a Gaussian filter. The reason why in the experiments we observe a Gaussian profile is because even if two photons are emitted within the polariton lifetime, the filter can induce an additional delay due to the longer lifetime a photon can survive within the filter. Capturing these features requires a more sophisticated  time-dependent treatment of light propagation through the filter that goes beyond the scope of this paper. In our simulations, we then restrict ourselves to a simplified model where the filter enters only as a number defining the probability of transmitting a photon, and it is, therefore, instantaneous. To obtain $\gtwo_{\rm sim}(0)$, we fit $\gtwo(\tau)$ with:
\be
f(\tau) = 1 + y_0 e^{-|\tau|/\tau_{\rm LP}}\,.
\label{eq:fit_simulation_g2}
\ee
with $\tau_{\rm LP}$ the polariton lifetime used in the simulations and $y_0$ an offset parameter. An example of this fit is shown in \Fig{fig:histograms_example}(b) in dashed line. The $\gtwo_{\rm sim}(0)$ is then calculated as 1+$y_0$. We note that with these simulations we can not extract the $\gtwo(0)$ directly from the histograms, as the Montecarlo method does not allow to have two events occurring within the same ${\rm d}t$. It is rather the fit of the exponential tails that allows to obtain $\gtwo(0)$. In \Fig{fig:Montecarlo_vs_Analytical} we compare the results of the Montecarlo simulations with the analytical expression given in \Eq{eq:sim_g2}, and find excellent agreement, therefore validating the analytical approach. \\

\subsection{Influence of the excitonic reservoir dynamics}
\label{subsecreservoir_dynamics}
In all the previous discussion, we have neglected the dynamics of the exciton reservoir. In the following, we discuss the correlations between the reservoir exciton and polariton populations.  
\subsubsection{Coupling mechanism}
Variations in $\nr$, due to fluctuations, translate into a global energy shift of the polariton spectrum via the reservoir exciton-polariton interaction $g_{\rm r}$. In formula:
\be
\omega_{n,\nr} = \omega_n + g_{\rm r}\nr\, .
\label{eq:reservoir_energy_fluct}
\ee
To understand the implications, let's remind that $G_{\freqF}^{\gF}(\omega) = G_{0}^{\gF}(\omega-\freqF)$ and $L_{\omega_n}^{\gsim}(\omega) = L_{0}^{\gsim}(\omega-\omega_n)$. Inserting these expressions in \Eq{eq:photon_transmission}, we obtain:
\be
\begin{split}
P_{n,\nr}^{\rm F} &=  \int_{-\infty}^{+\infty} {d}\omega\, L^{\gsim}_{0}(\omega-\omega_{n,\nr}) \times G_{0}^{\gF}(\omega-\freqF)\\ 
&= \int_{-\infty}^{+\infty} {d}\omega\, G_{0}^{\gF}(\omega) \times L^{\gsim}_{0}(\omega-\omega_{n,\nr} +\freqF).
\end{split}
\label{eq:prob_with_fluct}
\ee
where $P_{n,\nr}^{\rm F}$ is the modified photon transmission function which now depends on $\nr$ via to $\omega_{n,\nr}$. From the above expression, it is clear that the effect of changing the filter position or of a fluctuation of the reservoir exciton number is equivalent and corresponds to a horizontal displacement along the dispersive curve. 

\subsubsection{Influence of the reservoir-polariton correlations}
\label{subsubsec:reservor-polariton correlations}
The transfer of excitations from the reservoir to the polariton system correlates the statistics of $n$ and $\nr$. This is captured by a joint rate equations for the probability distribution $p(n,\nr)$:
\begin{widetext}
\be
\begin{split}
	\frac{d}{dt}p(n,\nr) &= \gsim \left[(n+1)p(n+1,\nr) -np(n,\nr)\right] \\
	&+\gr \left[n(\nr+1)p(n-1,\nr+1) -(n-1)\nr p(n,\nr)\right] \\
	&+ F \left[p(n,\nr-1) - p(n,\nr)\right] + \gd \left[(\nr+1)p(n,\nr+1) -\nr p(n,\nr)\right] \, .
\end{split}
\label{eq:joint_rate_eq}
\ee
\end{widetext}
where the first line describes the polariton dissipation, the second line the transfer of excitations from the reservoir to the polariton system and the last line the pump of reservoir exciton and the reservoir's other loss channels. Using the steady state solution of \Eq{eq:joint_rate_eq} $\proboccnr$, we can calculate $\gtwo(0)$ using \Eq{eq:sim_g2}, with the substitutions: $\sum_{n}\to \sum_{n,\nr}, P^{\rm F}_n\to P^{\rm F}_{n,\nr}$ and $\probocc \to \proboccnr$. In formula: 
\be
\begin{split}
	\gtwo(0) = \frac{\sum_{n,\nr} P_{n,\nr}^{\rm F} P_{n-1,\nr}^{\rm F} n (n-1) \proboccnr}{\sum_{n,\nr} \left( P_{n,\nr}^{\rm F} n \proboccnr \right)^2}\, .
\end{split}
\label{eq:sim_g2_nr}
\ee

\subsubsection{Polariton emission spectrum}
The emission spectrum of polaritons $S(\omega)$ can be obtained from
\be
S(\omega) = {\rm Re}\int d\tau e^{-i\omega \tau} \langle \creat (\tau) \annh \rangle \,.
\ee
To compute the correlation function $\langle \creat (\tau) \annh \rangle$, we use the evolution for the operator $\annh$ as derived from the master equation
\be
\frac{d}{dt}\annh = -\left[i \left( \omega_{n} + g_{\rm r} \right) + \frac{1}{2}\left(\gsim - \gr \nr\right)\right]\annh{\large {\tiny }}
\ee
Consequently one finds:
\be
S(\omega) = \langle n \rangle \frac{\left(\gsim - \gr \nr\right)/2}{\left( \omega_n - \omega\right)^2 + \left( \gsim - \gr \nr\right)^2}
\ee
Interestingly, the effective linewidth is $(\gamma - \gr \nr)$, due to the interplay between polariton loss via the cavity and polariton gain via the reservoir relaxation. 

\subsection{Effective interaction induced by the multiexcitons}
The coupling of polariton-modes to bi- and tri-exciton complexes is responsible for additional shifts in the transition frequency, effectively modifying the polariton-polariton interaction strength. We first infer the form of the corrections due to two-body collisions generating biexcitons, assuming a coupling Hamiltonian
\bea
H_\text{\rm PB} = g_{\rm PB}\vert \Cx\vert^2  \left[\hat B^\dagger \hat b^2 +  \hat b^{\dagger 2} \hat B\right]\, ,
\eea
where $\hat B$ ($\hat B^\dagger$) is the biexciton annihilation (creation) operator. We further assume that the dynamics of the biexciton mode is dominated by its non-radiative losses and can be eliminated adiabatically. This is true in the rotating frame given by the Hamiltonian $E_{\rm LP}(\hat b^\dagger \hat b +2 \hat B^\dagger \hat B)$, where all fast oscillations are removed. The biexciton mode dynamics is then given by a Heisenberg equation of the form:
\bea
\dot{\hat B} = -(i(\epsilon_{\rm B}-2 E_{\rm LP})+\gamma_{\rm B})\hat B -i g_{\rm PB}\hat b^2.
\eea
Assuming $\gamma_{\rm B} \gg \vert \epsilon_{\rm B}-2 E_{\rm LP}\vert$, we replace operator $\hat B$ by its steady-state value $-ig_{\rm PB}\vert \Cx\vert^2 \hat b^2/(i(\epsilon_{\rm B}-2 E_{\rm LP})+\gamma_{\rm B})$. Once re-injected in $H_{\rm PB}$, this result yields an effective polariton-polariton coupling:
\bea
H_\text{\rm PB} \simeq -2g_{\rm PB}^2\vert \Cx\vert^4 \frac{(\epsilon_{\rm B}-2 E_{\rm LP})}{(\epsilon_{\rm B}-2 E_{\rm LP})^2+\gamma_{\rm B}^2}b^{\dagger 2}b^2.
\eea
This interaction term contributes to renormalizing the value of $g$ by an amount $2g_{\rm PB}^2 \vert \Cx\vert^4\frac{(2 E_{\rm LP}-\epsilon_{\rm B})}{(\epsilon_{\rm B}-2 E_{\rm LP})^2+\gamma_{\rm B}^2}$ which was also predicted in \cite{TakemuraPRB17} and that we integrated in the definition of $\alpha_2$ in the main text.

Via the same approach, we estimate the impact of coupling to the triexciton via three-body collisions:
\bea
H_{\rm PT} =  g_{\rm PT}\vert \Cx\vert^3 \hat t^\dagger \hat b^3 + {\rm c.c.} \, ,
\eea 
with c.c. indicating the complex conjugate. We have introduced the triexciton mode $t$. Adiabatic elimination of the tri-exciton dynamics (assumed to be ruled by free triexciton oscillation at frequency $\epsilon_{\rm T}/\hbar$, non-radiative losses with rate $\gamma_{\rm T}$ and coupling to the polaritons via $H_{\rm PT}$), we obtain $\hat t \simeq \frac{-ig_{\rm PT}}{i(\epsilon_{\rm T}-3E_{\rm LP})}\hat b^3$, yielding to an effective polariton-polariton coupling due to three-body interaction
\bea
H_{\rm PT} & \simeq&  -2g_{\rm PT}^2\vert \Cx\vert^6 \frac{(\epsilon_{\rm T}-3 E_{\rm LP})}{(\epsilon_{\rm T}-3 E_{\rm LP})^2+\gamma_{\rm T}^2}\hat b^{\dagger 3}\hat b^3, \nonumber\\
\label{eq:3bodyPP}
\eea

We now use that:
\bea
\hat b^{\dagger 3}\hat b^2 = \hat n(\hat n-1)(\hat n-2)
\eea 
with $\hat n = \hat b^\dagger  b$. Finally, the frequency shift associated with Hamiltonian \eqref{eq:3bodyPP} is
\bea
\delta\omega_3 &=& g' (n-1)(n-2),\nonumber\\
\eea
with
\bea
g' = -2g_{\rm PT}^2\vert \Cx\vert^6 \frac{(\epsilon_{\rm T}-3 E_{\rm LP})}{(\epsilon_{\rm T}-3 E_{\rm LP})^2+\gamma_{\rm T}^2}.
\eea

To be more exhaustive, we have investigated two more phenomena that can affect these $3$-body shifts. First, triexcitons can be generated via another coupling mechanism which is the collision of a biexciton and a polariton. This is captured by an additional coupling Hamiltonian
\bea 
H_\text{PT}' = g'_\text{PT} \vert c_X\vert^3 \hat t^\dagger \hat b \hat B + {\rm c.c.}\, .
\eea
Second, the coupling constant $g_\text{PT}$ and $g_\text{PT}'$ might in principle be complex. Taking into account these two effects leads to a new expression for the  coefficient $g'$ appearing in $\delta\omega_3$ and $\delta\omega_3'$. We obtain:
\begin{widetext}
	\bea
	g'
	&=& 2 \vert g_\text{PT}\vert^2 \vert c_X\vert^{6}\vert f_T\vert^2\Delta_T\nonumber\\
	&+& \bigg\{ 4\vert c_X\vert^7\Delta_T \vert g_\text{PB}\vert^2 (\Delta_B\text{Re}g_\text{PT}' -\gamma_B \text{Im}g_\text{PT}' )+2\vert c_X\vert^{14} \Delta_T\vert g_{PT}' \vert^2 \vert g_\text{PB}\vert^4 \nonumber\\
	&+&2\text{Re}\Big[-(i\Delta_T+\gamma_T)(i\Delta_B+\gamma_B)(i\vert c_X\vert^{8}g_\text{PB}^* g_\text{PT}'g_\text{PT}  + \vert c_X\vert^{16}\vert g_\text{PB}\vert^2 (g_\text{PT}')^2  \nu^2 f_B)\Big]\bigg\} \vert f_B\vert^2 \vert f_T\vert^2 \,,
	\label{eq:triexciton_corrections}
	\eea
\end{widetext}
where we have denoted $\Delta_T = 3E_\text{LP}-\epsilon_\text{T}$, $\Delta_\text{B} = 2E_\text{LP}-E_\text{B}$, $f_{B,T} = [i\Delta_{B,T}+\gamma_{B,T}]^{-1}$. The first line corresponds to the dominant term which involves the square modulus of the polariton-triexciton coupling constant. The additional coupling $H'_\text{PT}$ induces a correction whose behavior across resonance with the biexction and triexciton frequencies is much more complex and depends on the phase of the couplings. This correction always involves the resonance factor of both the triexciton and the biexciton, i.e. the product $\vert f_T\vert^2\vert f_B\vert^2$ and higher powers of the Hopfiled coefficient $\vert c_X\vert$. For the analysis shown in \Fig{fig:g2_resonances} of the main text, we used only the dominant contribution expressed by the first line of \Eq{eq:triexciton_corrections}. The investigation of the other terms goes beyond the purpose of this work.


\subsection{Comparison with the experiment}
\label{subsubsec:comparison with experiment}

\subsubsection{Curves versus filter detuning}
\label{sec:curve_vs_filter_detuning}
\begin{figure}[]
	\includegraphics[width=0.9\columnwidth]{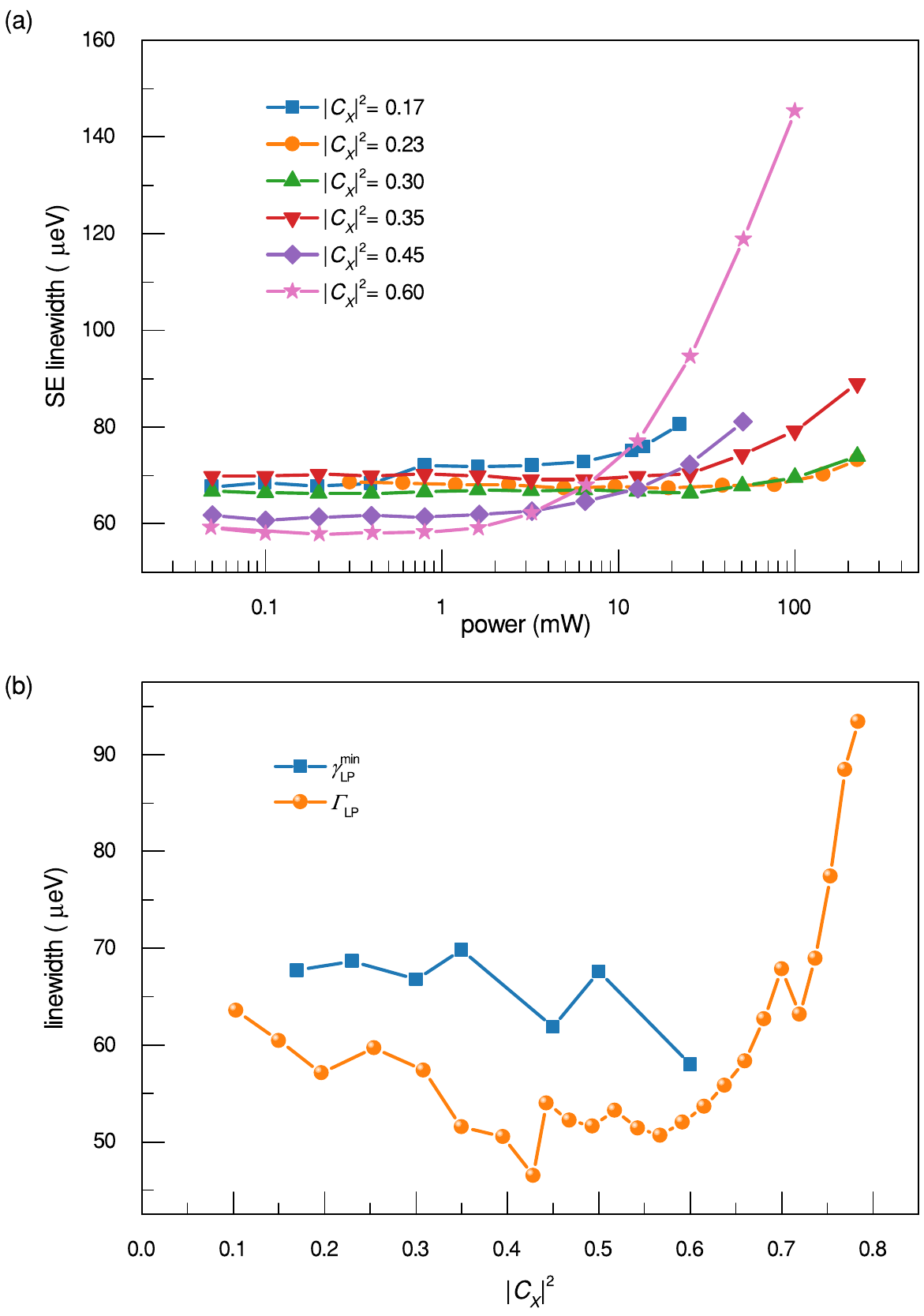}
	\caption{(a)\,SE linewidth versus power for different excitonic content. (b)\,Experimental linewidths of the resonant transmission ($\glp$) and of the polariton SE at low power $(\glpemmin)$, as a function of the exciton Hopfield coefficient. $(\glpemmin)$ is the value of $\glpem$ for the minimum measured excitation power, extracted from (a).}
	\label{fig:linewidth_res_vs_PL}
\end{figure}
To apply this theory to our set of measurements, we need an estimate of the exciton-polariton interaction constant $g_{\rm r}$. By considering that in resonant transmission the effect of the off-resonance exciton reservoir is highly suppressed, we can estimate the term $g_{r} \snr$ as the difference between the experimental linewidths $\glp$ (measured in resonant transmission) and $\glpem$ (measured in the SE regime), where $\snr$ is the standard deviation of reservoir exciton number calculated from the probability distribution $\proboccnr$. In \Fig{fig:linewidth_res_vs_PL}(a) we show the linewidth $\glpem$ as a function of the excitation power, for different excitonic content. From these measurements, we extract $\glpemmin$ as the value of the linewdith at the lowest measured power. \Fig{fig:linewidth_res_vs_PL}(b) shows the measured $\glp$ and $\glpemmin$ as a function of the exciton Hopfield coefficient. Using these measurements, we estimate the average value $\langle \glpemmin-\glp \rangle = 10.35 \pm 4.27\,\mu$eV. To model the data at $|\Cx|^2=0.42$ (\Fig{fig:Results}(c),\,(d)), we fixed $g_{r} \snr$ to this value. In GaAs, the off-resonant reservoir relaxation time $\tau_{\rm r}$ into the polariton system is typically on the order of few hundreds of picoseconds\,\cite{WaldherrNC18}. Here we fix it to a realistic value $\tau_{\rm r}=350$\,ps, resulting in $\gr \approx 1.9\,\mu$eV. Both $F$ and $\gd$ are unknown. However, given that the exciton reservoir is pumping not only the mode under investigation, but also the cross-polarized polariton mode with the rate $\gr$, we assume $\gd=\gr$. This leaves only $F$ and the polariton-polariton interaction strength $g$ as free parameters, while the other parameters are fixed by the experimental conditions. In particular, we use $\glpem=66.6\pm 0.2\,\mu$eV, which corresponds to a filter linewdith of $\gF=0.35\glpem$. We find good agreement with the data using $F=0.006\,{\rm ps}^{-1}$ and $g=0.04\,\glpem \approx 2.7\,\mu$eV. From the model, we estimate an average polariton number $\overline{n} \approx 0.03$, an average reservoir exciton number $\nrav \approx 1$ and the fluctuations $\snr \approx 1$,  which corresponds to $g_{\rm r}\approx 10\,\mu$eV. This results in the model shown in \Fig{fig:Results}(d) of the main text as a solid line. The corresponding model without including fluctuations (shown in \Fig{fig:Results}(d) as dashed line) can be obtained using \Eq{eq:sim_g2} with the same parameters and fixing $\nr=1$.
%
\begin{figure}[]
	\includegraphics[width=0.9\columnwidth]{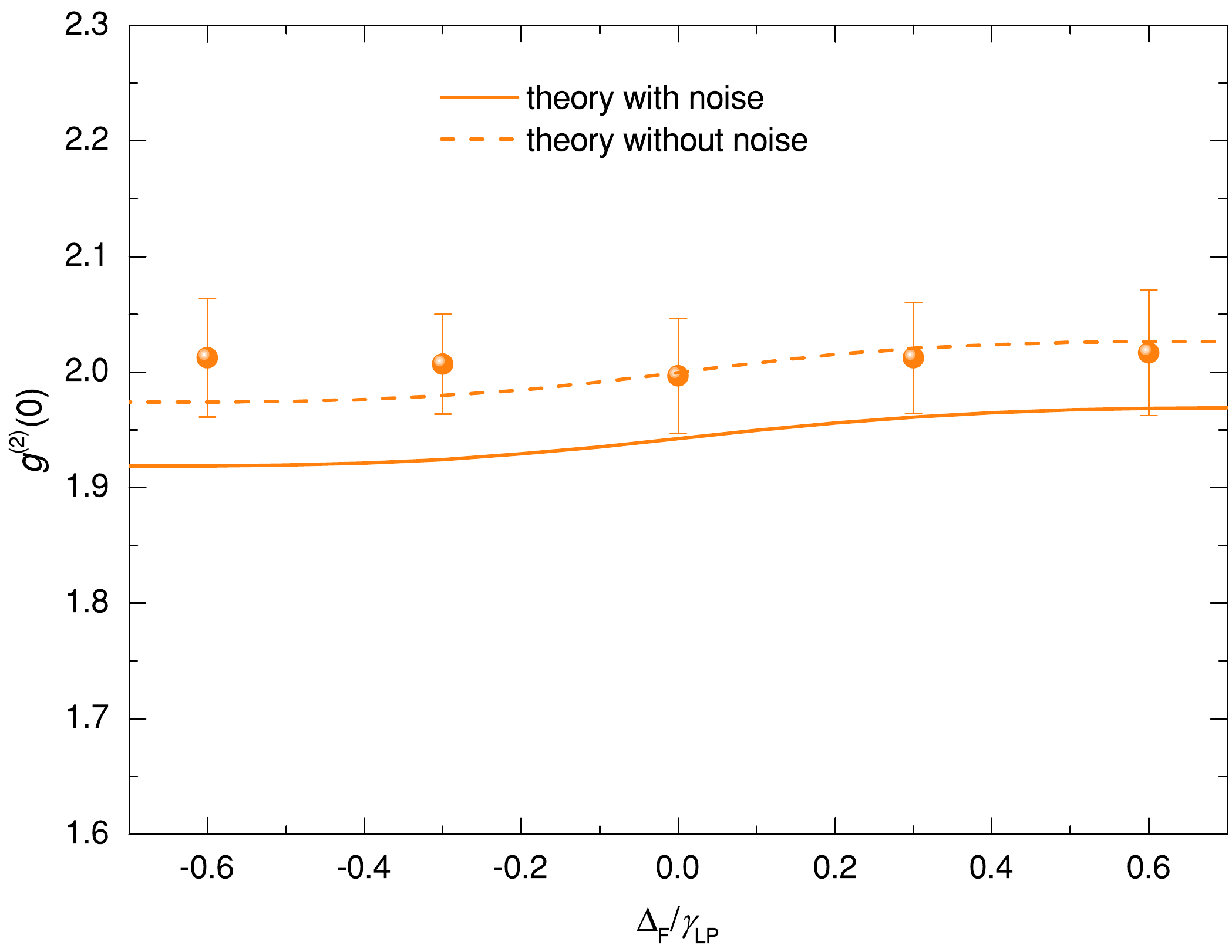}
	\caption{Model for the data at $|\Cx|^2=0.17$. Orange solid (dashed) line: model obtained by rescaling $g$ as expected from exciton-exciton interactions, with (without) including reservoir noise. The data points are the experimental data as in \Fig{fig:Results}(b).}
	\label{fig:Theory_vs_g _Cx_low}
\end{figure}
As it is possible to observe, the effect of fluctuations is to pull down the ``S-shaped'' curve as a result of the non-thermal character of the steady state distribution $\proboccnr$.

To model the data at $|\Cx|^2=0.17$, we use the same parameter, but with $g=0$. We note that in the simplified picture where polaritons interact only via exciton-exciton exchange interaction, at $|\Cx|^2=0.17$ we expect $g$ to be a factor $f=(0.17/0.42)^2$ smaller than the nonlinearity at $|\Cx|^2=0.42$. In \Fig{fig:Theory_vs_g _Cx_low} we show the result of the model, including the reservoir noise, in orange solid line, together with the experimental data as in the main text. As it is possible to see, a weak dispersive shape is predicted from the model, which slightly deviates from the experimental data at negative filter detuning. A better matching is obtained without including the reservoir noise, given in dashed line. Although this model matches the experimental data within the errorbars, in the main text we kept $g=0$ for simplicity. Importantly, this measurement shows significantly smaller nonlinearities than the corresponding measurement at $|\Cx|=0.42$. 

We used a similar reasoning for the model with and without noise shown in the inset of \Fig{fig:g2_resonances}, but this time we included both two-body (via $g$) and three-body (via $\gprime$) interactions.

Given the results described above, few points are worth mentioning. The data in \Fig{fig:Results}(b) in the main text are nicely described by the theoretical model without including the reservoir noise and using $g=0$ (dashed line). Interestingly, the data at higher excitonic content (\Fig{fig:Results}(d)) are better described by the noise model with $g=0.04\,\glpem \approx 2.7\,\mu$eV (solid line), while the model without noise is slightly offset towards higher values. This observation is consistent with the expectation that at larger excitonic content the polaritons are more affected by the interaction with the reservoir. On the other hand, for the data in the inset of \Fig{fig:g2_resonances}, both the model with and without noise seem to reproduce reasonably well the observed trend. We note that the data shown \Fig{fig:Results} and the one shown in \Fig{fig:g2_resonances} were acquired in two different cool-down sessions, meaning that the QW-microcavity system was warmed up to room temperature and cooled down again, at interval of few months, during which the sample was kept in vacuum. During this time, the experimental conditions might have changed slightly, including the sample position. Therefore, one has to be careful when comparing the two datasets, as the effect of the reservoir noise could be different due to different experimental conditions. A systematic study of the reservoir noise is not the purpose of the present work. Furthermore, it would require measuring curves versus filter detuning for different powers and at different cavity-exciton detuning during the same cool-down session. This would require a time $>$ 1 month of continuous measurements, which is not possible with the current setup that allows for only 4 weeks before helium is fully evaporated. Therefore, this will be material for follow up work. 

Finally, it is worth mentioning that from our model, one could argue that the reservoir noise is not significantly affecting the measured dispersive shape. This fact is quite surprising. In \Sec{sec:effect_of_noise} we provide additional quantification. 

\subsubsection{Curve versus filter cavity-exciton detuning}
\label{sec:vs_cavity_detuning}
Here we describe the details for the model shown in \Fig{fig:g2_resonances}.
Once $g=(\aone + \atwo)/2$ and $\gprime$ are calculated using \Eq{eq:alphas} and \Eq{eq:alpha3} respectively, we normalise them by the experimental linewidth $\glp$ measured in resonant transmission (shown in \Fig{fig:characterisation}(a)) to take into account losses which are not included in the model. As in the experiment, the filter position is fixed to $+0.6\glpem$, where $\glpem$ is the measured SE linewidth, and the filter width is kept fixed to 23\,$\mu$eV.  We then use these expressions in \Eq{eq:sim_g2}, using $\nr = 1$, with the energy shift calculated as $\omega_n=\omega_{\rm LP} + g(n-1)$ in the case where only two-body interactions are included, and as $\omega_n=\omega_{\rm LP} + g(n-1) + \gprime(n-1)(n-2)$ when also three-body interactions are included.
 
\subsubsection{Effect of increasing reservoir fluctuations}
\begin{figure}[]
	\includegraphics[width=0.9\columnwidth]{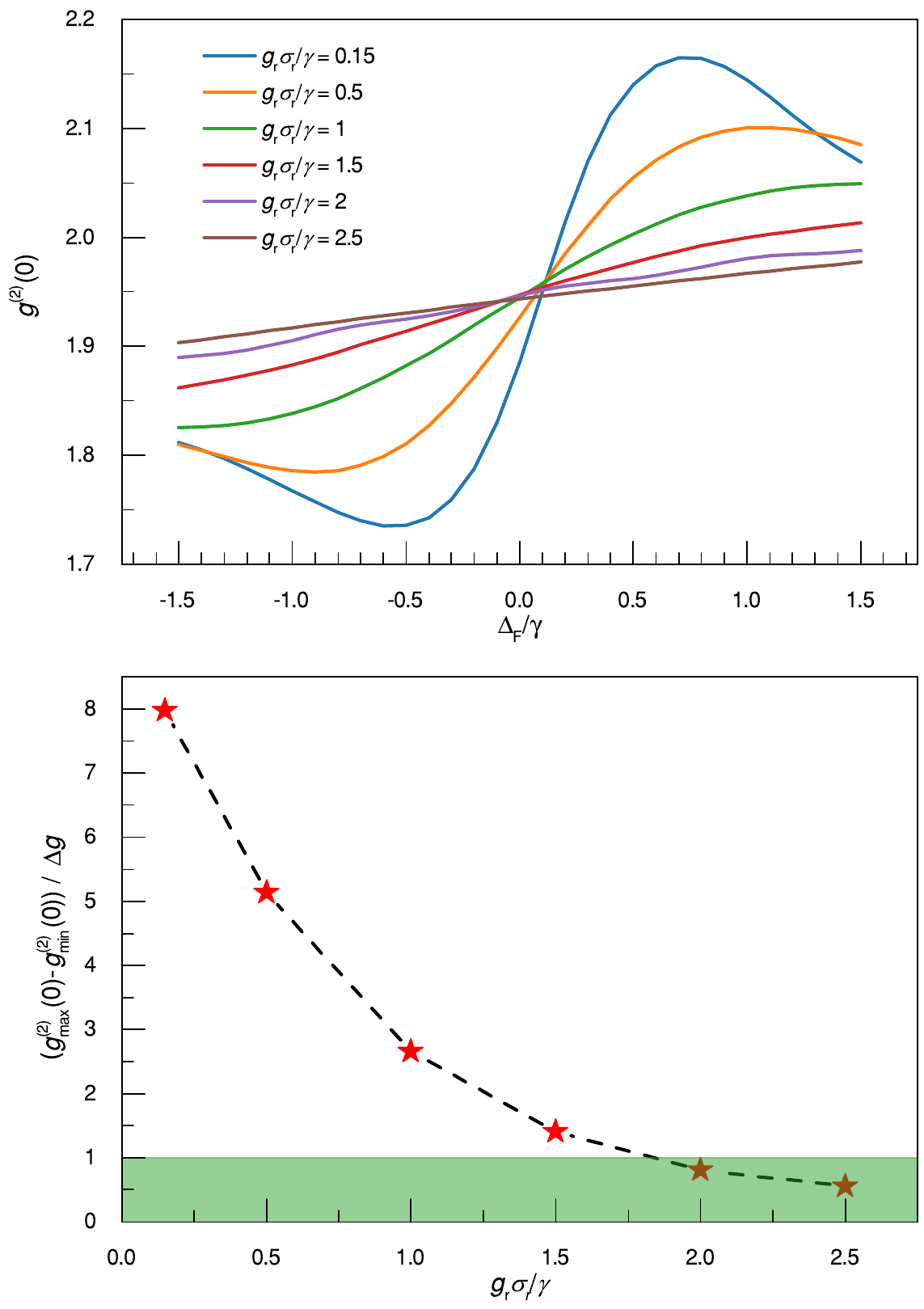}
	\caption{(a)\,$\gtwo(0)$ versus filter detuning for different values of $g_{\rm r}\snr/\gamma$ as indicated in the legend, where $g_{\rm r}$ was fixed to $10\,\mu$eV. (b)\,The difference between the maximum and minimum $\gtwo(0)$ values, calculated in the filter detuning range $\Df=(-0.6,0.6)\gamma$, and divided by the experimental error $\Delta g$, as a function of $g_{\rm r}\snr/\gamma$. The green band represents the region within the experimental error, for which the dispersive shape would not be resolved.}
	\label{fig:effect_of_noise}
\end{figure}
\label{sec:effect_of_noise}
As mentioned in \Sec{subsubsec:comparison with experiment}, we need to include reservoir-polariton correlations to properly describe the experimental data at higher excitonic content. Due to these correlations, described in \Sec{subsubsec:reservor-polariton correlations}, we expect the dispersive shape of $\gtwo(0)$ versus filter detuning to be completely washed out as the fluctuations of the exciton reservoir number increases above a certain threshold. It is, therefore, important to quantify this threshold, which is the purpose of this section. Starting from the steady state probability distribution $\proboccnr$ obtained with a certain set of parameters and with the fluctuations set to the experimental value $g_{\rm r}\snr / \gamma=0.15$, we manually increase $g_{\rm r}$ to amplify the effect of fluctuations and calculate the $\gtwo(0)$ as a function of filter detuning. For these calculations, we set $F=0.03\,$ps$^{-1}$ to have $\nrav\approx 5.3$, while all the remaining parameters are set as specified in \Sec{subsubsec:comparison with experiment}. In \Fig{fig:effect_of_noise}(a) we show the results of these calculations. We can clearly see that, as a result of the increased flcutuations, the dispersive shape is gradually washed out. To establish a more quantitative criteria, we calculate the difference between the maximum and minimum $\gtwo(0)$ ($g^{(2)}_{\rm max}(0)$ and $g^{(2)}_{\rm min}(0)$ respectively) within the range of filter detuning $(-0.6, 0.6)/\gamma$, which is the range used in the experiments. In \Fig{fig:effect_of_noise}(b) we summarise the results. As it is possible to observe, for $g_{\rm r}\snr / \gamma \geq 2$, the dispersive shape falls within the experimental error $\Delta g$ (green band), meaning it can not be resolved. Importantly, this means that our experimental measurement,  for which $g_{\rm r}\snr / \gamma=0.15$, are not significantly affected by the reservoir fluctuations.   

\end{appendix}

\end{document}